\documentclass{aa}  

\usepackage{graphicx}
\usepackage{color}
\usepackage{soul}
\usepackage{txfonts}
\usepackage{hyperref}
\newcommand{\aelem}{$\alpha$-elements}
\newcommand{\aelemsing}{$\alpha$-element}
\newcommand{\feh}{{\rm [Fe/H]}}
\newcommand{\afe}{{\rm [\alpha/Fe]}}
\newcommand{\teff}{T_{\rm eff}}
\newcommand{\logg}{\log{g}}
\newcommand{\msun}{M$_\odot$}

\newcommand{\aesopus}{\AE SOPUS}
\newcommand{\abcoeff}{\left\{a, b\right\}^{\psi}_{j,i,r}}

\newcommand{\aevol}{\afe_{\rm ev}}

\begin{document}

   \title{From spectroscopic abundances to evolutionary $\afe$ in stellar models}
   

   \author{Pedro Diaz Reeve
          \inst{1, 2}
          \and
          Aldo Serenelli \inst{1, 2}
          }

   \institute{Institute of Space Sciences (ICE, CSIC), Carrer de Can Magrans S/N, E-08193, Cerdanyola del Vallès, Spain
         \and
             Institut d’Estudis Espacials de Catalunya (IEEC), Carrer Esteve Terradas, 1, Edifici RDIT, Campus PMT-UPC, E-08860, Castelldefels, Spain\\
             \email{diaz@ice.csic.es}
             }

   \date{Received; accepted}

 
  \abstract
   {Large-scale, high-resolution spectroscopic surveys map detailed chemical abundance patterns for up to millions of stars. The richness of this information is used in a wide range of studies, from galactic archaeology and stellar populations to processes such as planet engulfment by host stars. However, libraries of stellar evolution models and isochrones still rely on a simplistic description of chemical abundance patterns: all stars have either a solar-scaled or, at most, a constant enrichment for all \aelem. Moreover, the definition of $\alpha$-enrichment is survey-dependent, and it is usually not the most relevant quantity for stellar structure and evolution; spectroscopic surveys and stellar models do not speak the same language.
   In the era of precision stellar astrophysics, it is necessary to go beyond this simplified approach to have stellar evolution models that are on par with the wealth of data coming from these spectroscopic surveys to enable more accurate determination of fundamental stellar parameters.}
   {We aim to develop a perturbative calculation method that allows to compute stellar evolution tracks for stars with general patterns of $\alpha$-elements. We also aim to devise a simple method that, using the observed abundances of individual $\alpha$-elements, allows the computation of [$\alpha$/Fe] that correctly reflects the true impact of observed abundance patterns on stellar models rather than relying on a spectroscopically defined $\alpha$-enhancement whose definition and adopted element sets often vary between surveys.}
   {We have quantified the response of stellar evolutionary tracks to variations of individual \aelem\ through simple polynomial expansions. For a general \aelem\ abundance pattern, these individual contributions can be linearly combined to produce a synthetic evolutionary track based on the exact observed abundance pattern. As a second step, we present a method to determine the so-called evolutionary $\alpha$-enhancement ($\aevol$), i.e. a constant enhancement for all \aelem\ that encodes the true impact of the observed pattern on stellar models. This allows us to mimic the traditional $\alpha$-enhancement widely used in the available libraries of stellar models and isochrones.}
   {We validate our method for computing synthetic tracks and deriving $\aevol$ using a set of target stars with detailed APOGEE information. We then quantify the difference between the spectroscopic $\afe$ and $\aevol$ for different surveys, and we present a minimal list of chemical elements required for accurate calculation of stellar models. Finally, we present \texttt{ALCHEMY}, a simple and fast tool that allows the user to compute $\aevol$ given a general \aelem\ pattern.}
   {}


   \maketitle
%

\section{Introduction}

Chemical composition plays a key role in stellar structure and evolution, and hence determines the main properties of stars, such as their luminosities, ages, and effective temperatures. Stars form through the accretion of gas and dust from the interstellar medium (ISM), which is chemically enriched by the nucleosynthetic yields from previous stellar populations dispersed via stellar winds and supernova explosions (SNe; see, e.g., \citealp{2011ApJ...734...48L,2016MNRAS.456.1235S,2020SSRv..216...68G}.In this context, Core Collapse Supernovae (CCSNe) are the main producers of \aelem, i.e. elements that are produced by the progressive addition of $\alpha$ ($^4$He) particles to $^{12}$C, which correspond to the most abundant isotopes of O, Ne, Mg, Si, S, Ca, Ar and Ti (e.g. \cite{1995ApJS..101..181W}; \cite{2006NuPhA.777..424N}). In contrast, Type Ia supernovae predominantly release Fe-peak elements, although they also contribute significantly to the production of intermediate-mass elements (\cite{2003fthp.conf..331T}). The effect of \aelem\ on the evolution of low-mass stars has long been recognized \citep{1993ApJ...414..580S,2000ApJ...532..430V}. Large sets of stellar evolution tracks and isochrones are routinely computed in two flavors: assuming a solar-scaled distribution of metals, or an $\alpha$-enhanced (or depleted) one in which all \aelem\ are scaled by a common factor \citep{2008ApJS..178...89D,2014ApJ...794...72V,2021ApJ...908..102P,2024MNRAS.527.2065P}. 

Large-scale spectroscopic stellar surveys have become commonplace over the last two decades. Surveys such as the Apache Point Observatory Galactic Evolution Experiment (APOGEE; \citealp{2017AJ....154...94M}) in the infrared range, Galactic Archaeology with HERMES (GALAH; \citealp{2014IAUS..298..322A,2015MNRAS.449.2604D}), the Large Sky Area Multi-Object Fiber Spectroscopic Telescope (LAMOST; \citealp{2012RAA....12..735D, 2012RAA....12..723Z}) and the Gaia-ESO Survey (GES; \citealp{2022A&A...666A.121R}) in the visible range routinely provide rich information on the chemical makeup of stars. The oncoming ESO's 4MOST survey will do the same for millions of stars in both high- and low-resolution 
(\citealp[respectively]{2019Msngr.175...35B,2019Msngr.175...30C}). 
These surveys provide standard spectroscopically determined quantities such as effective temperature ($\teff$), surface gravity ($\logg$), 
and $\feh$\footnote{The standard spectroscopic bracket notation is adopted throughout this work such that for any two chemical elements X and Y with number densities ${\rm N_X}$ and ${\rm N_Y}$ respectively, $[X/Y] = \log{\left( {\rm N_{X}/N_{Y}} \right)_*} - \log{\left( {\rm N_{X}/N_{Y}} \right)_\odot}$, where the first term refers to the star and the second to the Sun.}, along with, crucially, individual abundances for a wide range of chemical elements, and sometimes a global determination of $\alpha$-enhancement ($\afe$). 

Studies published over the last two decades have explored the effects of individually modifying the abundances of the most relevant elements from their scaled-solar values on stellar evolutionary models. \citet{2007ApJ...666..403D} found that, for models computed at fixed metallicity ($Z$), enhancing the abundances of C, N, O, and Ne produces hotter and more luminous evolutionary tracks with shorter stellar lifetimes. In contrast, increasing the abundances of Mg, Si, S, Ca, Ti, and Fe results in cooler, less luminous models with longer lifetimes. Similar qualitative behavior was reported by \citet{2022MNRAS.511.3198W} for models in which the abundances of C, N, O, Mg, Si, and Fe were varied individually. In addition to their effects on the evolutionary tracks, individual abundance variations also modify stellar lifetimes. \citet{2007ApJ...666..403D} showed that changes in the abundances of these elements typically alter main-sequence lifetimes by approximately 5$\%$. Although such differences are smaller than the uncertainties associated with many traditional age-dating techniques, such as standard isochrone fitting for field stars, they are comparable to the precision targeted by modern asteroseismic analyses and high-quality isochrone fitting. As stellar age determinations continue to approach the 5-10\% level of precision, systematic effects arising from detailed chemical abundance patterns become increasingly important.

An alternative approach was adopted by \citet{2012ApJ...755...15V}, who investigated the effects of enhancing the abundances of C, N, O, Ne, Na, Mg, Si, S, Ca, and Ti by +0.40 dex while keeping [Fe/H], rather than $Z$, fixed. Although enhancing a single element at fixed $Z$ requires reducing the abundances of the remaining metals, whereas fixing [Fe/H] leads to a change in the total metal mass fraction, they found the same overall qualitative behavior as in previous studies. This is because the [Fe/H] values corresponding to the enhanced mixtures differ by less than 0.05 dex from the reference composition for all elements except oxygen, implying that calculations performed at fixed $Z$ provide a good approximation to those obtained when a single metal is enhanced at constant [Fe/H].

Beyond their impact on stellar evolutionary tracks and lifetimes, variations in the abundances of $\alpha$ elements also modify the internal stellar structure and, consequently, the predicted asteroseismic properties of stellar models. Recent studies (e.g., \citet{lindsay2026effectdifferentmethodsaccounting}) have shown that accounting for these abundance variations is important for the precise modeling of oscillating stars and the determination of their fundamental properties.

Although the response of stellar evolution models to changes in individual elements can be easily modeled and understood, the implementation of detailed chemical composition in general-purpose large-scale libraries of stellar evolutionary tracks and isochrones is computationally expensive, as each element introduces an additional dimension to the parameter space over which models are computed (see e.g., \citet{2023ApJS..268...29S}). In reference to \aelem, this has been traditionally circumvented by using $\alpha-$scaled evolutionary tracks. However, this presents two limitations. First, \aelem\ in a star do not all have the same fractional variation with respect to Fe. Second, \aelem\ that are more easily observable in spectroscopic surveys are not necessarily the most relevant for stellar modeling, and vice-versa. Notable examples include Ti and Ca, which are relatively straightforward to measure spectroscopically but have a limited impact on stellar structure and evolution, primarily due to their low overall abundances. On the other hand O, for example, has a rather large impact on stellar modeling but is more difficult to determine spectroscopically. As a result, the \textit{spectroscopic} $\alpha-$enhancement will typically be different from the one that is relevant for stellar structure and evolution, so a direct mapping between spectroscopic and stellar modeling $\afe$ is not correct. 

This work has two main goals: to develop a methodology to compute stellar evolution models that can take advantage of the rich information provided by spectroscopic surveys for individual stars, without increasing the dimensionality of the parameter space such that the calculation of large-scale libraries of evolutionary models becomes impractical; and to present a simple method that allows mapping any set of spectroscopically observed \aelem\ into an $\alpha-$enhancement value that best represents their impact on stellar evolutionary models, therefore allowing a more rigorous use of the widely available $\alpha-$scaled evolutionary tracks.

The article is organized as follows: Sect.~\ref{s:abundances} sets the scene from an observational perspective, the scope of our work in terms of types of stars (FGK and red giants), presents the grid of stellar models computed for this work, and our analysis on the impact of \aelem\ in stellar evolution models. In Sect.~\ref{s:alphaevol} we develop and test our method to compute synthetic tracks for general \aelem\ abundance patterns and establish the concept of evolutionary $\alpha$-enhancement, $\aevol$. Sect.~\ref{s:alchemy} presents \texttt{ALCHEMY}, a \texttt{python} tool to compute $\aevol$ for any general $\alpha$-element pattern. Sect.~\ref{s:results} compares $\aevol$ with $\afe$ determined by three large-scale spectroscopic surveys and presents a minimal list of \aelem\ abundances that are required for a physically accurate determination of $\aevol$. Finally, Sect.~\ref{s:conclusion} summarizes this work and delineates future possibilities.

\section{\aelem\ in the sky and in stellar models}
\label{s:abundances}

\subsection{General context on chemical abundances}\label{ss:surveys}

The primary motivation for this work is given by the immense progress in observational stellar astronomy over the past two decades, particularly due to large-scale spectroscopic surveys that provide enormous amounts of high-quality stellar spectra. In combination with improvements and automation of analysis techniques, as well as with the development of more realistic stellar atmosphere models and radiative transfer modeling, these surveys provide individual elemental abundances of many chemical elements for hundreds of thousands of stars. 

\begin{figure*}
  \centering
  \includegraphics[width=\textwidth]{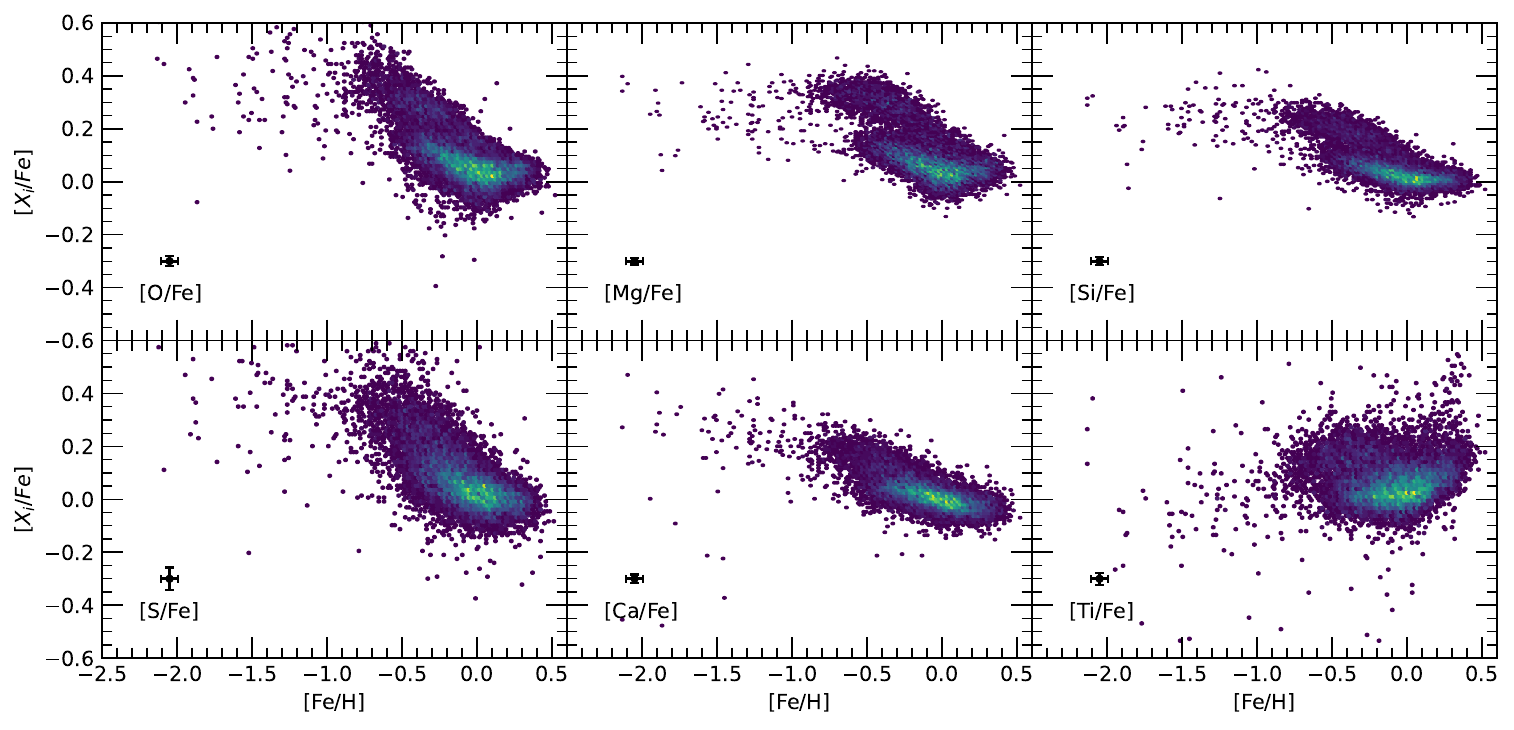}
  \caption{Observational spectroscopic chemical abundance ratios [$X_i$/Fe] vs. [Fe/H] for $X_i$ = O, Mg, Si, S, Ca and Ti, for stars included in the APOKASC-3.}
  \label{f:alpha_1}
\end{figure*}

In this work, our interest focuses on \aelem. Figure~\ref{f:alpha_1} shows the abundance ratios relative to Fe of \aelem\ of stars in the APOKASC-3 catalog \citep{2025ApJS..276...69P}. While it is clear that the abundances of \aelem\ show strong correlations among each other, their relation to Fe is different for all elements. These variations among elements justify the importance of treating them individually, not through global scaling as traditionally done in stellar modeling.  

\subsection{Stellar evolution models}\label{ss:models}

Stellar evolution models were computed with the GARching STellar Evolution Code \texttt{GARSTEC} version 20.1 \citep{2008Ap&SS.316...99W}. The models for this work employ OP radiative opacities \citep{2005MNRAS.360..458B}, complemented at low temperatures by molecular and dust opacities computed with the \AE SOPUS web interface\footnote{ Available at: \href{https://stev.oapd.inaf.it/cgi-bin/aesopus}{https://stev.oapd.inaf.it/cgi-bin/aesopus}} (\cite{2009A&A...508.1539M}; \cite{2022ApJ...940..129M}) and conductive opacities from Potekhin as updated in \cite{2007ApJ...661.1094C}. Nuclear reaction rates for hydrogen burning are adopted from Solar Fusion II \citep{2011RvMP...83..195A}, and helium-burning rates from NACRE \citep{Angulo1999ACO}. The nuclear reaction network implemented in GARSTEC explicitly follows $p$, $^{3}$He, $^{4}$He, $^{12}$C, $^{13}$C, $^{14}$N, $^{15}$N, $^{16}$O, and $^{17}$O nuclei during H-burning. For He-burning and subsequent burning phases, the network includes $p$, $n$, $^{4}$He, $^{12}$C, $^{16}$O, $^{20}$Ne, $^{24}$Mg, $^{28}$Si, and $^{56}$Ni. For the equation of state we adopted the FreeEOS \citep{2012ascl.soft11002I}, and stellar atmospheres are modeled using the VAL-C $T - \tau$ relation from \cite{1981ApJS...45..635V}. Convection is treated within the framework of mixing length theory following \cite{1968pss..book.....C} and the mixing length parameter is calibrated using a standard solar model which, in combination with the adopted atmospheric boundary conditions, low-temperature opacities, $T-\tau$ relation and the solar mixture leads to a $\alpha_{\rm MLT}= 2.0530$, which is adopted throughout all the models in this work. The corresponding initial solar composition is $X_{\odot,{\rm ini}} = 0.70988$, $Y_{\odot,{\rm ini}} = 0.27190 $ and $Z_{\odot,{\rm ini}} = 0.01822$.

Gravitational settling is included for models with masses below 1.35 $M_{\odot}$ following the formalism of \cite{1994ApJ...421..828T}. To account for the reduced efficiency of atomic diffusion in more massive stars, the settling efficiency is gradually suppressed between 1.25 and 1.35 $M_{\odot}$ through a linear scaling factor that increases from 0 to 1 across this interval, and diffusion is completely neglected above 1.35 $M_{\odot}$. This approximation has only a minor effect on the structure of RGB stars. Additional turbulent mixing beneath the convective envelope is modeled as a diffusive process using the prescription of \cite{2012ApJ...755...15V}. Convective boundary mixing is treated using the diffusive overshooting formalism of \cite{1996A&A...313..497F}, adopting a constant efficiency parameter of $f=0.02$ at all convective boundaries. To prevent unrealistically large overshooting regions in stars with very small convective cores, the overshooting efficiency in the core is reduced linearly from its standard value to zero for models with masses between 1.40 and 1.00 $M_{\odot}$. The reference solar chemical composition is taken from \cite{2022A&A...661A.140M}, hereafter MB22. Accordingly, a solar metal-to-hydrogen ratio of $(Z/X)_{\odot}=0.02249$ is adopted to define the spectroscopic abundance scale, corresponding to [Fe/H]$=0.00$.


To compute the initial composition of models with solar-scaled mixture for any \feh, we use a linear chemical enrichment law defined as 
\begin{equation}\label{eq:y-z}
    Y =  Y_{\text{SBBN}} + \Delta Z, 
\end{equation}
and the basic relation
\begin{equation}\label{eq:zfeh}
\left(\frac{Z}{X}\right) = \left(\frac{Z_{\odot,{\rm ini}}}{X_{\odot,{\rm ini}}}\right) \times 10^{\rm{[Fe/H]}} ,
\end{equation}
that is applicable as long as the mixture of metals is solar.  Above, 
\begin{equation}\label{eq:delta}
\Delta = \frac{Y_{\odot,{\rm ini}} - Y_{\text{SBBN}}}{Z_{\odot,{\rm ini}}},
\end{equation}
and $\text{Y}_{\text{SBBN}} = 0.2485$ is the helium mass fraction resulting from Standard Big Bang Nucleosynthesis \citep{2011ApJS..192...18K}.

From these relations and the identity $X+Y+Z=1$, we obtain
\begin{equation} \label{eq:zfinal}
Z = \frac{\left(1 - Y_{\text{SBBN}}\right)}{1+\Delta + 10^{-\text{[Fe/H]}}/\left(\frac{Z_{\odot,{\rm ini}}}{X_{\odot,{\rm ini}}}\right)}.
\end{equation}

In order to quantify the effects of \aelem, in particular O, Ne, Mg, Si, S, Ca and Ti, on stellar evolutionary tracks, we computed dedicated Rosseland mean opacities and stellar evolutionary models in which these elements have been varied one at the time by $\pm$0.20, $\pm$0.40 and $\pm$0.60~dex relative to the scaled-solar mixture, while keeping all other chemical abundances fixed at scaled-solar values. For these models, the initial composition is computed as above, with the only difference that, in equation~\ref{eq:zfeh}, the ratio of solar 
$\left(Z_{\odot,{\rm ini}}/X_{\odot,{\rm ini}}\right)$ is replaced by 
\begin{equation}\label{eq:zvaried}
    \left( \frac{Z}{X} \right)_{i} = \left(\frac{Z_{\odot,{\rm ini}}}{X_{\odot,{\rm ini}}}\right) + \frac{A_{i} \cdot 10^{\varepsilon_{i}}}{A_H \cdot 10^{\varepsilon_H}} \left(10^{\delta_i} - 1 \right); \quad i\text{ = O, Ne, ..., Ti}. 
\end{equation}
Here $A_H$ and $A_i$ are the atomic masses of hydrogen and element $i$ respectively, $\varepsilon_i$ is the MB22 abundance of element $i$ on the scale where $\varepsilon_H=12$, and $\delta_i$ is the variation of the abundance of element $i$ with respect to the value of MB22 in dex. $X$ and $Y$ are then computed using the enrichment law and the normalization relation. For each of the varied abundances we have used the modified metal mixture and consistent opacity tables.

For this work, the grid of stellar models spans the ranges 0.60 to 2.00 $M_{\odot}$ for stellar mass, with a 0.10 $M_{\odot}$ step, and [Fe/H] from -2.60 to 0.60 in increments of 0.20 for [Fe/H] $\leq$ 0.00, and 0.10 for [Fe/H] $>$ 0.00. These ranges encompass most scientific cases in which abundance patterns for \aelem\ different from a solar-scaled one may play a role in the evolution of stars with solar-like oscillations in the Milky Way. In this work, we focus on evolutionary phases from the main sequence (MS) to the red giant branch (RGB). We do not explicitly include red clump and asymptotic giant branch models in this work, although we expect our results for RGB stars will also hold for those evolutionary phases, as the external structure of those stars resembles that of the red bump and upper RGB, respectively.

Once the evolutionary tracks were computed, each of them was interpolated into a grid of equivalent evolutionary points (EEPs), as is usually performed in the calculation of isochrones \citep{1996yCat.6040....0G,2016ApJS..222....8D}. For this work, we have defined eight primary EEPs: the first three correspond to the main sequence, the transition between EEPs 3 and 4 encapsulates the sub-giant branch phase, and the remaining points encompass the RGB phase. Further details regarding the normalization process and the specific EEP definitions are provided in Appendix \ref{appendix a}.
Bringing  all evolutionary tracks to a standardized EEP grid makes it simple to quantify the impact of variations in chemical composition (Sect.~\ref{ss:cmpeffect}), as well as to create synthetic tracks for any composition (Sect.~\ref{s:alphaevol}).

\subsection{Impact of \aelem\  on stellar evolutionary models} \label{ss:cmpeffect}

\begin{figure}[!htb]
  \centering
  \includegraphics[width=\columnwidth]{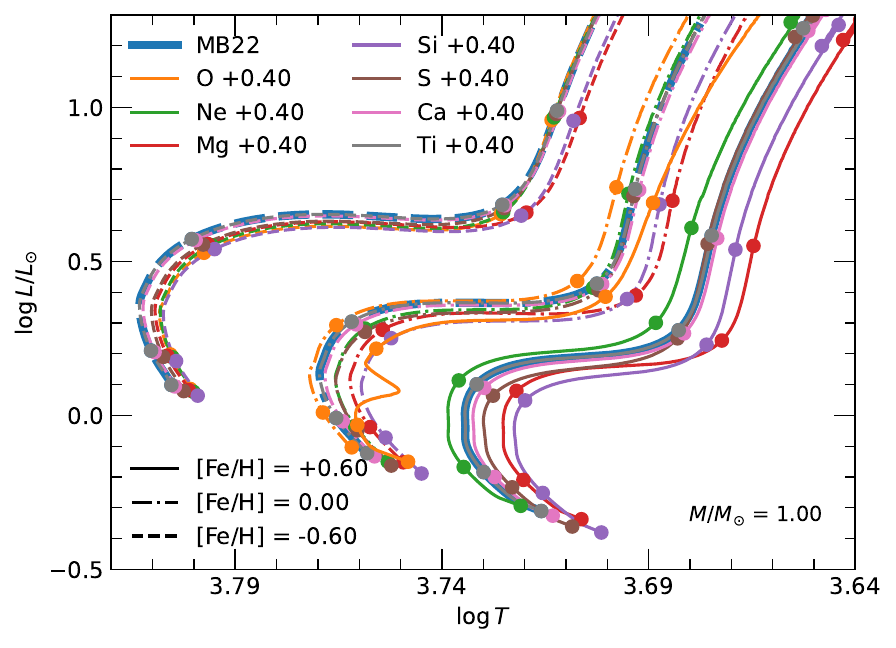}
  \caption{Hertzsprung-Russell diagram for stellar evolution tracks with scaled-solar chemical composition and for models in which the abundance of each $\alpha$-element has been individually enhanced by +0.40~dex from the scaled-solar value. Tracks are shown along the main-sequence, sub-giant branch and beginning of the red giant branch phases, for 1.00 $M_{\odot}$ and [Fe/H] = +0.60 (solid lines), 0.00 (dashed lines) and -0.60 (dotted lines). Points show EEPs 1 to 5 defined in Appendix \ref{appendix a} are shown.}
  \label{f:mstracks}
\end{figure}

\begin{figure}[!htb]
  \centering
  \includegraphics[width=\columnwidth]{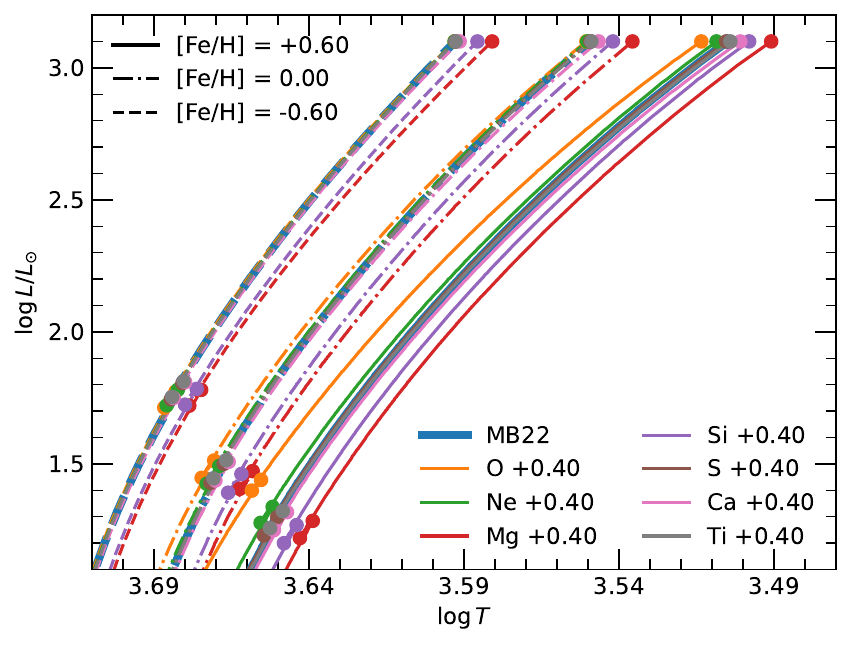}
  \caption{Hertzsprung-Russell diagram for stellar evolution tracks with scaled-solar chemical composition and for models in which the abundance of each $\alpha$-element has been individually enhanced +0.40 dex from the scaled-solar value. Tracks are shown along the red giant branch phases, for 1.00 $M_{\odot}$ and [Fe/H] = +0.60 (solid lines), 0.00 (dashed lines) and -0.60 (dotted lines). Points show EEPs 6 to 8 defined in Appendix \ref{appendix a} are shown.}
  \label{f:rgbtracks}
\end{figure}

Figure~\ref{f:mstracks} illustrates the impact of increasing the abundance of \aelem\ individually by +0.40~dex, on the MS, the sub-giant branch (SGB) and the lower RGB. Similarly, Fig.~\ref{f:rgbtracks} presents the effects along the red giant branch up to $\log L/L_{\odot} = 3.10$. These results are displayed for 1.00~\msun\ models with [Fe/H]= +0.60, 0.00, and $-0.60$. To complement the analysis and provide additional context, Table~\ref{t:fractions} summarizes the initial mass fractions of helium, and heavy elements for each stellar model with individually altered $\alpha$-element abundances, analyzed in Figs.~\ref{f:mstracks} and \ref{f:rgbtracks}.

\begin{table*}
\centering          
\begin{tabular}{c c c c c c c}     
\hline\hline       
    &  \multicolumn{2}{c}{[Fe/H]=$-0.60$} & \multicolumn{2}{c}{[Fe/H]=0.00} & \multicolumn{2}{c}{[Fe/H]=+0.60} \\ 
    \hline
      & $Y_i$ & $Z_i$ & $Y_i$ & $Z_i$ & $Y_i$ & $Z_i$ \\
     \hline
     MB22 & 0.25388 & 0.00419 & 0.26915 & 0.01608 & 0.32024 & 0.05586 \\
     $[$O/Fe$]$=+0.40 & 0.25719 & 0.00677 & 0.28112 & 0.02540 & 0.35406 & 0.08219 \\
     $[$Ne/Fe$]$=+0.40 & 0.25489 & 0.00498 & 0.27284 & 0.01896 & 0.33121 & 0.06440 \\
     $[$Mg/Fe$]$=+0.40 & 0.25419 & 0.00443 & 0.27027 & 0.01695 & 0.32363 & 0.05850 \\
     $[$Si/Fe$]$=+0.40 & 0.25427 & 0.00449 & 0.27057 & 0.01719 & 0.32453 & 0.05920 \\
     $[$S/Fe$]$=+0.40 & 0.25405 & 0.00432 & 0.26975 & 0.01655 & 0.32207 & 0.05729 \\
     $[$Ca/Fe$]$=+0.40 & 0.25392 & 0.00422 & 0.26927 & 0.01617 & 0.32061 & 0.05615 \\
     $[$Ti/Fe$]$=+0.40 & 0.25388 & 0.00419 & 0.26915 & 0.01608 & 0.32026 & 0.05588 \\
\hline                  
\end{tabular} 
\vspace*{0.2 cm}
\caption{Initial helium and heavy elements mass fractions, $Y_i$ and $Z_i$, used in stellar models with individually altered $\alpha$-element abundances, for models with [Fe/H] = -0.60, 0.00 and 0.60 dex.}
\label{t:fractions}      
\end{table*}

The qualitative response of evolutionary tracks in the MS and SGB can be understood in simple terms from homology relations for stars dominated by radiative transport, even though these stars have (thin) convective envelopes. The basic relation states that the luminosity of a star scales with the stellar mass, an average radiative opacity $\kappa$ and the mean molecular weight $\mu$ as
\begin{equation} \label{eq:basic}
L\propto \frac{M^3 \mu^4 }{\kappa}.
\end{equation}

When increasing the abundance of an \aelemsing\, the most direct effect is that the Rosseland mean opacity increases as well, leading to a decrease in the luminosity of the star. A similar homology relationship explains the decrease in effective temperature. This is most clearly seen in tracks with increased Mg and Si because their contribution to the Rosseland mean opacity, a combination of atomic physics and their abundance, is relatively large. The impact of S is smaller, but noticeable in the \feh=+0.60 model, while Ca and Ti have almost negligible impact due to their low abundance. 

For oxygen, the response of evolutionary tracks is more complex. An increase in the abundance of O leads to an increase in the Rosseland opacity. But it also leads to a rather large increase in the initial $Z$ of the models by Eq.~\ref{eq:zvaried} which, following Eq.~\ref{eq:y-z} results in a larger initial helium abundance and, consequently, a larger $\mu$ (see Table~\ref{t:fractions}). The latter induces changes opposite to those of an increased opacity by means of Eq.~\ref{eq:basic}. Although this is formally true for all metals, the increase in $\mu$ is strongly dominant only for O given its dominant contribution to the total metallicity, of which it accounts for more than 40\% for a solar mixture. This is clearly seen for the \feh$=+0.60$ model with $+0.40$~dex increase in [O/Fe], which is much hotter and more luminous than the reference model. At smaller \feh, the same fractional change in O leads to a smaller net effect on $\mu$ and $Y$, with the increase in Rosseland opacity almost compensating for the change $\mu$ at \feh$=0.00$ and dominating \feh~$< 0.00$. A qualitatively similar behavior is imprinted by changes in the Ne abundance, but substantially smaller due to its smaller contribution to the stellar metallicity. For other \aelem, the change they produce in $\mu$ is negligible and, to all practical matters, they impact stellar evolution only through their effect on the Rosseland opacity. 

Finally, in the case of the model with \feh$=+0.60$ model with [O/Fe]=$+0.40$~dex a distinct effect is the appearance of a convective core, as reflected in the morphology of the track. This is driven by the increase in the initial helium due to the relation between Z and Y that we adopt in our models. In our models, the appearance of the convective core at 1~\msun\ already occurs for [O/Fe]=$+0.20$~dex. The transition is mass and O-enhancement dependent. 

The behavior of models along the RGB mimics the discussion above, with the impact of the Rosseland mean at low temperatures, where the dominant effect is due to Mg and Si in that order. Along the RGB, changes in Mg, Si, S, Ca and Ti abundances affect the effective temperature scale in a rather [Fe/H] independent way. In our experiments, Ca and Ti, elements typically observed in spectroscopic surveys and used to establish the $\alpha$-enhancement of stars, have almost completely negligible impact on evolutionary tracks.  

Models with stellar masses in our range of interest show an analogous response to changes in individual \aelem\ as discussed here.


\section{Calculation of [$\alpha$/Fe] for stellar evolution models}\label{s:alphaevol}

\subsection{Towards synthetic evolutionary tracks: local behavior at EEPs}\label{ss:quadratic}

\begin{figure}[!htb]
  \centering
  \includegraphics[width=\columnwidth]{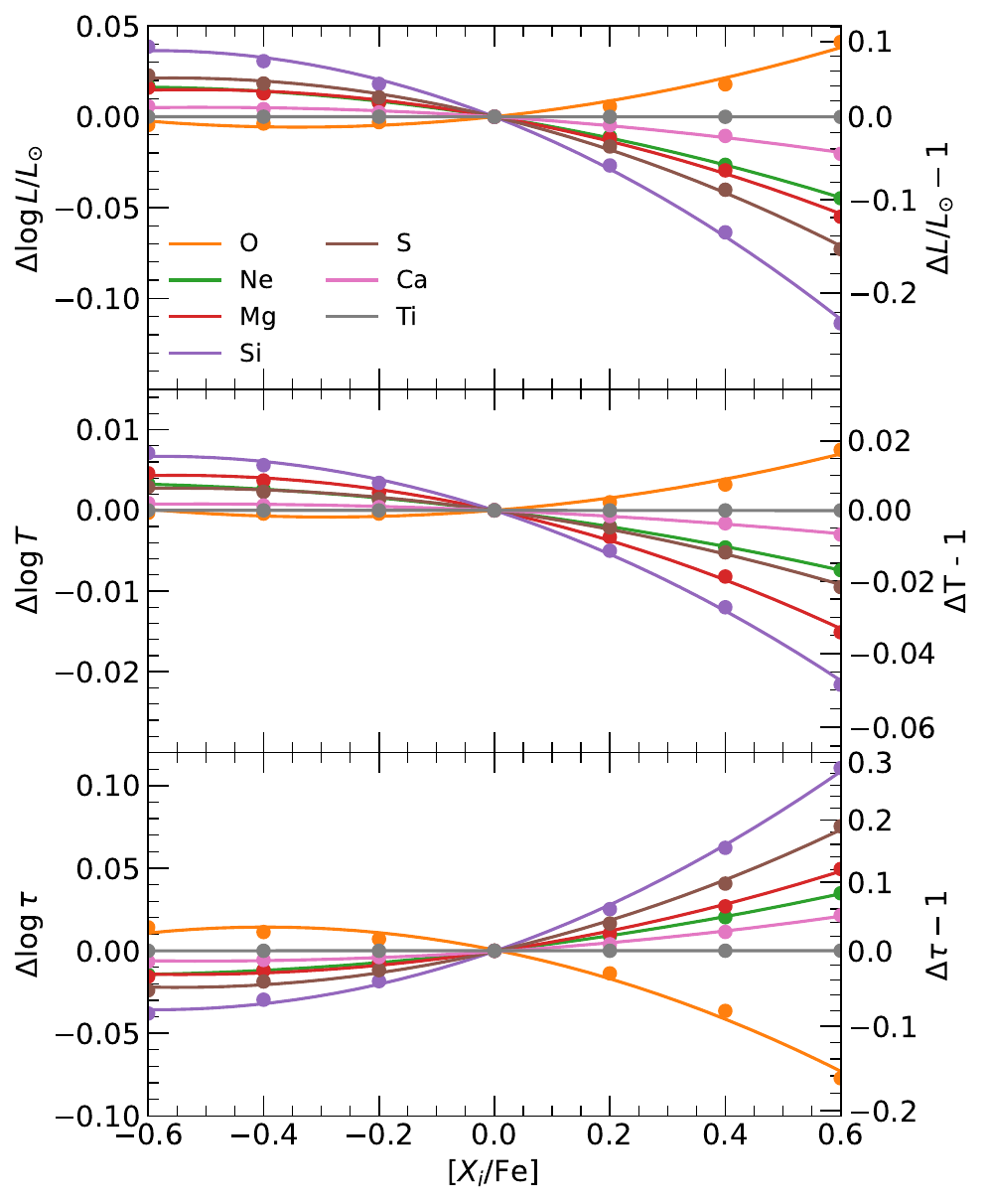}
  \caption{$\Delta \log L/L_{\odot}^{syn}$ (top), $\Delta \log T^{syn}$ (middle) and $\Delta \tau^{syn}$ (bottom) as a function of [$X_i$/Fe] for $X_i$ = O, Ne, Mg, Si, S, Ca and Ti. Dots show the value of $\psi_i - \psi_{r, i}$ at the second main EEP, identified as the point just before the central hydrogen abundance drops below 60$\%$ of its initial value (see \ref{appendix a} for details), for 1.00 $M_{\odot}$ and [Fe/H] = 0.00 model star, on the main-sequence and solid lines show the quadratic fit in eq. (\ref{eq:quadratic}) for each element.}
  \label{f:quadraticfit}
\end{figure}

We have analyzed the behavior of physical quantities of main interest, $\log L/L_{\odot}$, $\log T$, and $\log \tau$ relative to a reference solar-scaled track of given mass and [Fe/H] as a function of the enhancement of each $\alpha$-element. We have found that the variation of these quantities can be described locally, i.e. at a given EEP, by a quadratic function of the variation in the abundance of the $\alpha$-element. As an example, Figure~\ref{f:quadraticfit} shows the changes of these quantities in the second main EEP as a function of the enhancement of \aelem, and the quadratic functions that best model the variations. 

Therefore, in each EEP $j$, the difference of any physical quantity $\psi$ with respect to its value on the reference evolutionary track $r$, when the abundance of the $\alpha$-element ${\rm X_i}$ is varied, can be written as follows
\begin{equation}\label{eq:quadratic}
\Delta \psi_{j,i,r} = a^{\psi}_{j,i,r} [{\rm X_i/Fe}]^2 + b^{\psi}_{j,i,r} [{\rm X_i/Fe}].
\end{equation}
Here, $[{\rm X_i/Fe}]$ is the enhancement of the $\alpha$-element 
and $a^{\psi}_{j,i,r}$ and $b^{\psi}_{j,i,r}$ are the coefficients of the quadratic function. The coefficients are computed using the evolutionary tracks described in \ref{ss:models}, as illustrated in Figure~\ref{f:quadraticfit}. 

We have computed the quadratic functions for all the 300 reference stellar tracks defined by our grid of stellar masses and [Fe/H] values described in Sect.~\ref{ss:models}. For each reference track, we store the coefficients for $\log L/L_{\odot}$, $\log T$, and $\log \tau$, at each EEP, and for each $\alpha$-element. When considered as functions of stellar mass or [Fe/H], the behavior of the fitting coefficients is smooth, which allows interpolation of the grid to any mass or [Fe/H] value. We denote by $\abcoeff$ the complete set of coefficients that define the interpolating functions defined in Eq.~\ref{eq:quadratic}.

\subsection{Synthetic evolutionary tracks: linear combination}\label{ss:synthetic_linear}

Results from the previous section can be used to compute synthetic evolutionary tracks, i.e. generated from the quadratic expansions, for a general \aelem\ abundance pattern. 

The simplest approach is to assume that the functional response of tracks with a given mass and [Fe/H] depends very little on the abundance of \aelem, i.e. that each $\alpha$-element can be treated independently. Therefore, for any $\alpha$-element abundance pattern, Equation~\ref{eq:quadratic} can be used to compute individual variations, and these can then be summed up to compute the total variation. At each EEP $j$, the physical quantity $\psi$ is computed as 
\begin{equation}\label{eq:synthetictrack}
    \psi^{syn}_{j,r} = \psi_{j,r} + \Delta \psi^{syn}_{j,r} = \psi_{j,r} + 
    \sum_{i}{\left(a^{\psi}_{j,i,r} [{\rm X_i/Fe}]^2 + b^{\psi}_{j,i,r} [{\rm X_i/Fe}] \right)}.
\end{equation}
Here, $\psi_{j,r}$ is the value of $\psi$ in the reference track $r$ in point $j$. We denote by \textit{synthetic} the evolutionary tracks computed in this way.

\begin{table*}
\centering          
\begin{tabular}{c c c c c c c}     
\hline\hline       
 & Star 1 & Star 2 & Star 3 & Star 4 & Star 5 & Star 6   \\ 
\hline                    
   KIC & 10118942 & 2309595 & 9754284 & 5253542 & 6145937 & 7191496 \\  
   $m/M_{\odot}$ & $1.184^{+0.047}_{-0.046}$ & $1.167^{+0.084}_{-0.077}$ & $1.276^{+0.038}_{-0.064}$ & $1.077^{+0.053}_{-0.053}$ & $1.065^{+0.067}_{-0.066}$ & 1.017 $\pm$ 0.038 \\ 
   
   [Fe/H] & -0.539 $\pm$ 0.015 & -0.066 $\pm$ 0.007 & 0.077 $\pm$ 0.005 & 0.275 $\pm$ 0.004 & -0.186 $\pm$ 0.006 & -2.094 $\pm$ 0.012 \\
   
   [O/Fe] & 0.077 $\pm$ 0.149 & 0.090 $\pm$ 0.023 & 0.117 $\pm$ 0.087 & 0.146 $\pm$ 0.040 & 0.133 $\pm$ 0.042 & 0.604 $\pm$ 0.068 \\

   [Ne/Fe] & 0.077 $\pm$ 0.149 & 0.090 $\pm$ 0.023 & 0.117 $\pm$ 0.087 & 0.146 $\pm$ 0.040 & 0.133 $\pm$ 0.042 & 0.604 $\pm$ 0.068 \\


   [Mg/Fe] & 0.132 $\pm$ 0.024 & 0.124 $\pm$ 0.011 & -0.022 $\pm$ 0.012 & 0.018 $\pm$ 0.010 & -0.003 $\pm$ 0.012 & 0.342 $\pm$ 0.034 \\

   [Si/Fe] & 0.121 $\pm$ 0.027 & 0.025 $\pm$ 0.013 & 0.076 $\pm$ 0.013 & 0.081 $\pm$ 0.010 & -0.117 $\pm$ 0.014 & 0.291 $\pm$ 0.032 \\

   [S/Fe] & 0.170 $\pm$ 0.104 & 0.050 $\pm$ 0.031 & -0.008 $\pm$ 0.042 & -0.118 $\pm$ 0.034 & -0.313 $\pm$ 0.034 & 0.671 $\pm$ 0.177 \\

   [Ca/Fe] & -0.059 $\pm$ 0.033 & 0.051 $\pm$ 0.013 & 0.011 $\pm$ 0.012 & -0.055 $\pm$ 0.008 & 0.212 $\pm$ 0.015 & 0.270 $\pm$ 0.055 \\

   [Ti/Fe] & 0.192 $\pm$ 0.137 & 0.024 $\pm$ 0.022 & -0.166 $\pm$ 0.077 & -0.219 $\pm$ 0.046 & 0.231 $\pm$ 0.027 & 0.263 $\pm$ 0.055 \\ 

   \hline 



   [$\alpha$/Fe]$_1$ & 0.047 $\pm$ 0.022 & 0.048 $\pm$ 0.022 & -0.002 $\pm$ 0.022 & 0.006 $\pm$ 0.022 & -0.044 $\pm$ 0.022 &  0.356 $\pm$ 0.024 \\

   \hline

   [$\alpha$/Fe]$_2$ & 0.093 $\pm$ 0.036 & 0.052 $\pm$ 0.008 & 0.015 $\pm$ 0.014& -0.058 $\pm$ 0.031 & -0.055 $\pm$ 0.007 &  0.433 $\pm$ 0.039 \\

   [$\alpha$/Fe]$_3$ & 0.093 $\pm$ 0.031 & 0.052 $\pm$ 0.008 & 0.018 $\pm$ 0.011& -0.061 $\pm$ 0.035 & -0.057 $\pm$ 0.006 &  0.433 $\pm$ 0.039 \\   

\hline                  
\end{tabular} 
\vspace*{0.2 cm}
\caption{Abundance ratios for the $\alpha$-elements considered in this work for six stars contained in the APOKASC-3 catalog. Masses for stars 1 to 5 are extracted from (\cite{2017ApJS..233...23S}), and mass for star 6 is extracted from the APOKASC-3 catalog. [$\alpha$/Fe]$_1$ are spectroscopic [$\alpha$/Fe] values contained in the APOKASC-3 catalog. [$\alpha$/Fe]$_2$ and [$\alpha$/Fe]$_3$ are $\aevol$ ratios calculated using the general method and \texttt{ALCHEMY} respectively. Star 6 has no data for the mass so we assumed 1.00 $M_{\odot}$.
\label{t:stars}      
}
\end{table*}

\subsection{Synthetic evolutionary tracks: testing with Kepler stars}\label{ss:synthtest}

To test the accuracy of synthetic tracks, we selected six stars with known masses from the dwarfs and subgiants APOKASC catalog \citep{2017ApJS..233...23S}, and used their APOGEE DR17 \citep{2022ApJS..259...35A} chemical abundances. These stars  cover a broad range of [Fe/H] values and enhancements for all \aelem.  The relevant properties of these stars are provided in Table~\ref{t:stars}. Note that it is not our purpose here to model these stars. We simply take them as real-life test cases for comparing synthetic to full evolutionary tracks, from the Zero Age Main Sequence (ZAMS) to the tip of the Red Giant Branch (RGB-tip). 

For each of the six test cases, we computed stellar evolutionary models with GARSTEC using the reported mass and \aelem\ abundances,  with the MB22 as our reference solar-scaled composition pattern in all cases. Radiative opacities from OP and \aesopus\ were explicitly computed for the detailed composition of each star. For Ne, whose abundance cannot be determined spectroscopically for cool stars, we adopted the same enhancement of O, under the assumption that the galactic evolution of both elements follow each other closely, i.e. [Ne/Fe] = [O/Fe] in all cases. 

To compute the synthetic tracks, we used GARSTEC to compute the reference tracks for all stars with the mass and [Fe/H] values reported in Table~\ref{t:stars} using the solar-scaled composition pattern MB22. We then interpolated the set $\abcoeff$ in mass and [Fe/H] for each of the six stars and used Eq.~\ref{eq:synthetictrack} to compute the synthetic stars by linear addition. 

\begin{figure}[!htb]
  \centering
  \includegraphics[width=\columnwidth]{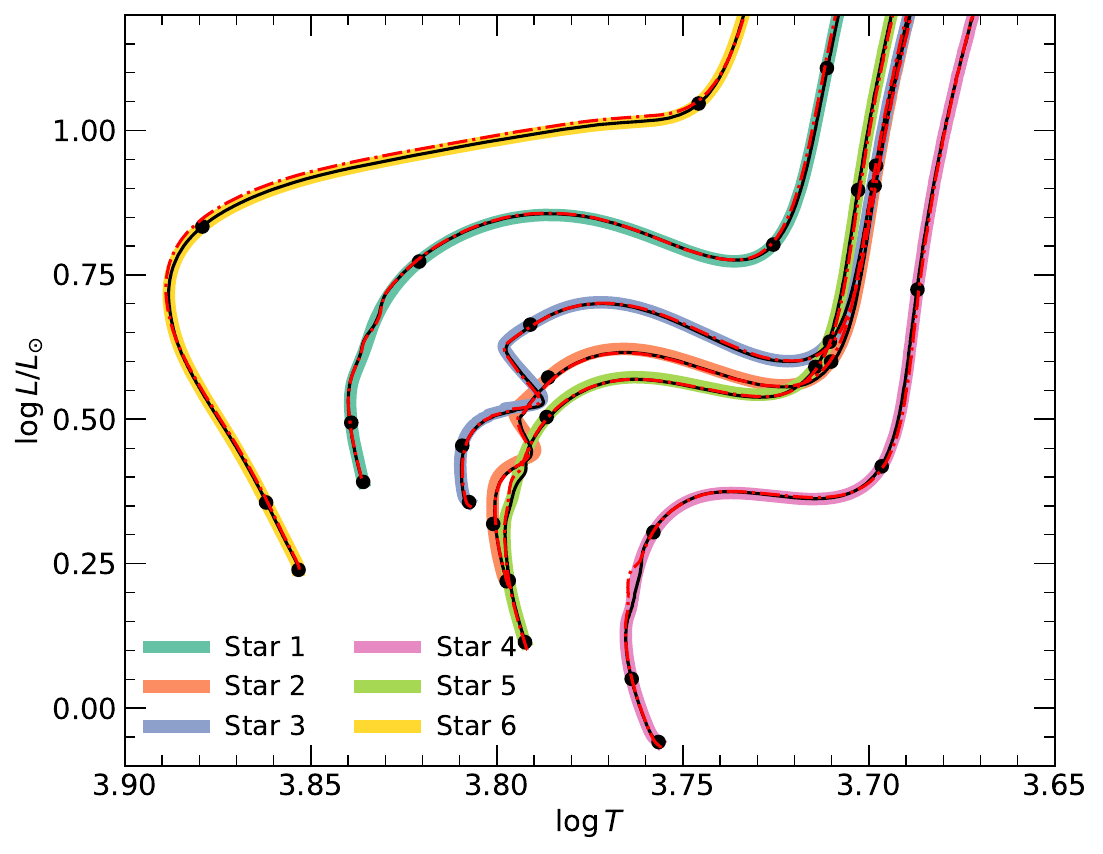}
  
  \includegraphics[width=0.95\columnwidth]{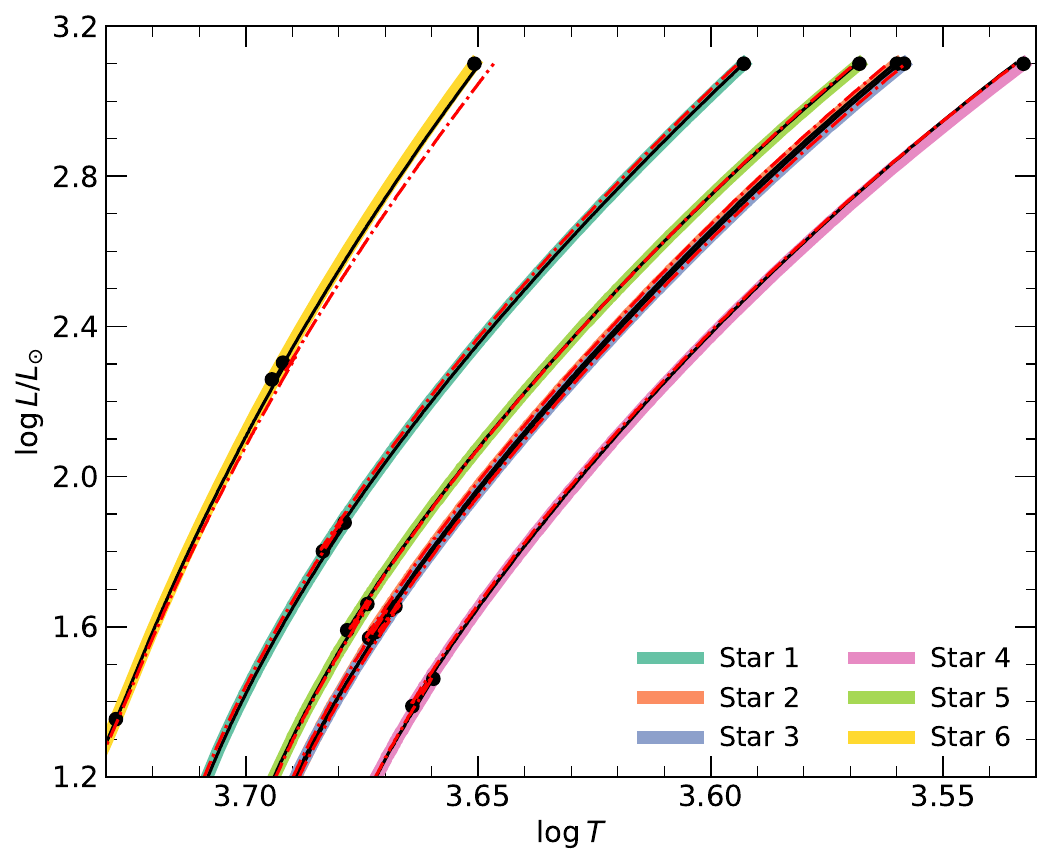}
  
  \caption{Top panel: main-sequence and sub-giant branch phase along the Hertzsprung-Russell diagram modeled using observational data for target stars 1 to 6 in Table \ref{t:stars}. Thick colored lines show the evolutionary tracks computed with GARSTEC and the detailed abundance pattern of each star. Black solid lines depict the synthetic tracks (Eq.~\ref{eq:synthetictrack}). Red dashed lines show synthetic tracks computed using $\aevol$. Bottom panel: same as above but for the red giant branch phase.
  \label{fig:tracks_stars}}
\end{figure} 

\begin{figure}[!htb]
  \centering
  \includegraphics[width=\columnwidth]{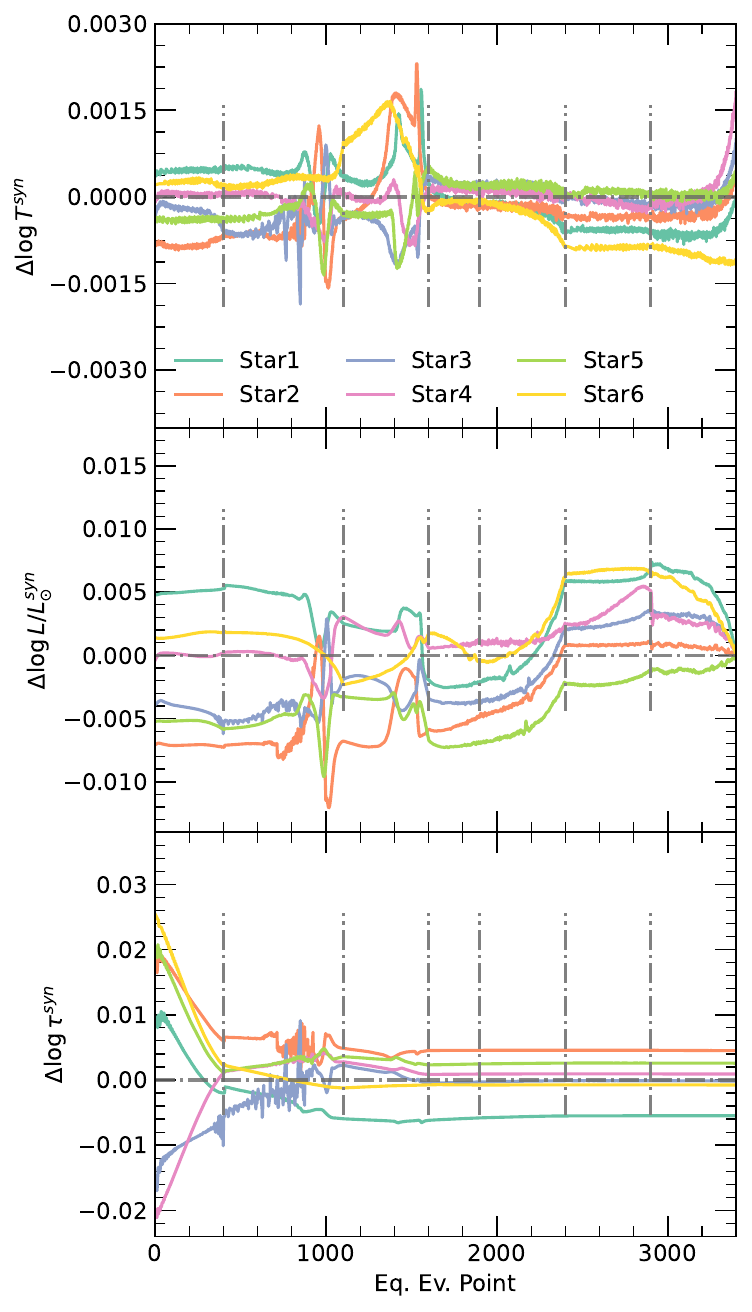}
  \caption{Differences between synthetic and full evolutionary tracks as a function of EEPs from the ZAMS to the RGB for the most relevant quantities, $\teff$, luminosity and age along the EEPs from the ZAMS to the RGB. Gray vertical lines show main EEPs 2 to 7. Points 1 and 8 are the limits 0 and 3400.
  \label{f:eeps_stars_syn} }
\end{figure}

Figure~\ref{fig:tracks_stars} compares the evolutionary tracks from the ZAMS to the RGB-tip. The agreement between the full evolutionary tracks and the synthetic tracks, shown in thick colored and black lines, respectively, is excellent for all six test cases. A more quantitative assessment is shown in Fig.~\ref{f:eeps_stars_syn}, where differences in effective temperature, luminosity, and age are shown, as a function of equivalent evolutionary points, for all six test cases. For the effective temperature, differences remain always smaller than 0.3\%, i.e. not more than 20 or 30~K, much smaller than realistic spectroscopic and photometric temperature determinations. For the luminosity, differences are typically smaller than 1\%, with the largest deviations reaching a 2\% level (Star 2 and Star 5) at the main sequence hook related to the development of small convective cores (see Fig.~\ref{fig:tracks_stars}).

The results in this section show that the response of stellar models to changes in the \aelem\ abundances can be accurately reproduced by a linear superposition of variations from individual elements, provided that a model with the same mass and [Fe/H] values (solar-scaled mixture) is used as reference. The need of a known mass and [Fe/H] might seem restrictive at first sight, but in fact it is common place as grids of stellar models in mass and [Fe/H] are routinely interpolated e.g. when building isochrones. The reference track can be, in fact, trivially computed from a grid of stellar evolution tracks such as ours.  

In summary, the set of coefficients $\abcoeff$ encodes the response of models to changes in \aelem\ abundances and, together with the usual interpolation methods to construct isochrones, can be used to compute synthetic evolutionary tracks for any arbitrary (within the range explored in this work) \aelem\ pattern. The method presented here offers for the first time, to the best of our knowledge, a simple method to move away from the "one-for-all" alpha-elements enhancement paradigm.

\subsection{Synthetic evolutionary tracks: $\aevol$ models}\label{ss:synthetic_alpha}

The procedure described in the previous section enables the use of an arbitrary \aelem\ abundance pattern to build individual evolutionary tracks. While the method is general and fast, as it requires only simple algebraic calculations, it is practical when the number of synthetic tracks that need to be computed is limited, e.g. when prior information on stellar mass is known, so that the total number of synthetic tracks needed is not too large. In a more general situation, e.g. when large samples of stars need to be analyzed and observational errors need to be accounted for as well, computing synthetic tracks for individual stars can become prohibitively expensive. Therefore, it is necessary to develop an alternative methodology that can be used in such cases. 

The simple answer is to use the traditional methodology of assuming that all \aelem\ have the same enhancement. The caveat with this approach is that spectroscopic surveys report $\alpha-$enhancements in a variety of ways that depend, among other things, on the spectral wavelength coverage that defines which elements are observable and the analysis methodology, e.g. global $\alpha-$enhancement determination or some sort of average using commonly observed \aelem. The latter, in particular, might not necessarily include elements that most affect the stellar evolution tracks, as discussed in previous sections. Stellar evolution and stellar spectroscopy do not necessarily mean the same when speaking of $\alpha-$enhancement, nor do spectroscopic surveys among themselves. Here, we develop a simple procedure that translates the spectroscopic determination of \aelem\ abundances into a unique $\alpha-$enhancement value appropriate for stellar evolutionary models. 

The linear expansion from previous sections can be used to build synthetic tracks that assume a constant enhancement for all \aelem, as
\begin{equation}\label{eq:alphatrack}
    \psi_{j,r}^{\alpha} = \psi_{r, j} + \Delta \psi_{j,r}^{\alpha} = \psi_{r, j} + \sum_i \left( a_{j, i,r}^{\psi} [\alpha/Fe]^2 + b_{j, i,r}^{\psi} [\alpha/Fe] \right), 
\end{equation}
where the difference with respect to Eq.~\ref{eq:synthetictrack} is that a unique $\afe$ value, so far undetermined, is used for all elements. We have shown in the previous section that synthetic tracks built using the linear expansions are accurate, i.e. they approximate  true evolutionary tracks computed with arbitrary \aelem\ abundance patterns very well. Then, for a physical quantity $\psi$, we define the residual at each equivalent evolutionary point $j$ as:
\begin{equation}\label{eq:epsilon}
    \varepsilon_{\psi, j} = \frac{\psi_j^{syn} - \psi_j^{\alpha}}{\psi_{\rm max} - \psi_{\rm min}} = \frac{\Delta \psi_j^{syn} - \Delta \psi_j^{\alpha}}{\psi_{\rm max} - \psi_{\rm min}}, 
\end{equation}
where we have dropped the index $r$ for simplicity. The normalization factor included in the definition accounts for the widely different ranges that physical quantities might have along the evolutionary track, bringing all of them to a comparable scale. These residuals can be used to construct a penalty function that globally measures how well a track built with a unique $\afe$ value describes the effects of a given \aelem\ abundance pattern. 

Upon examining different alternatives, we have found that a satisfactory penalty function is
\begin{equation}\label{eq:f_alpha}
F = \sum_{j} \Delta \tau_{j} \left[ \varepsilon_{\log{L},j}^2 + \varepsilon_{\log{T_{\rm eff}},j}^2\right],
\end{equation}
where $\Delta \tau_j$ is the evolutionary time step at point $j$.

The $\afe$ value that best approximates the true evolutionary track, which we assume to be the \textit{synthetic} track from Sect.~\ref{ss:synthetic_linear}, can be obtained by minimization of F: 
\begin{equation}
\frac{\partial F}{\partial \afe} = 0.
\end{equation}
The solution to this equation determines the stellar evolution \textit{best-fit} constant $\alpha$-enhancement value, $\aevol$, that best reproduces the impact of a given \aelem\ abundance pattern on stellar evolution tracks. Minimization of this quartic equation is carried out using the \texttt{minimize} function of the SciPy package \citep{2020SciPy-NMeth}.

We also calculate the uncertainty in $\aevol$ due to errors in the abundances. Assuming that errors in spectroscopic abundances are not correlated\footnote{This is a strong, and likely, wrong assumption, but spectroscopic surveys do not report covariance matrices for errors.}, we can propagate these errors individually. Therefore, for each element $i$, we build the functions $F_i^{\pm}$ as before, but with $[X_i$/Fe] of the element $i$ increased/decreased by its observational error, i.e.  $[X_i/{\rm Fe}] \rightarrow  [X_i/{\rm Fe}] \pm \sigma_i$. Minimization of $F_i^{\pm}$ then leads to the determination of 1-$\sigma$ uncertainties in 
$\aevol$ due to uncertainties in element $i$, i.e.
\begin{equation}
\sigma^{\pm}_{\afe,i} = \afe^{\pm}_{\rm ev,i} - \aevol.
\end{equation}
The total uncertainty in $\aevol$ is then computed by quadratic addition, 
\begin{equation}
\sigma_{\aevol}^2 = \sum_i \left( \frac{|\sigma^+_{\afe,i}| + |\sigma^-_{\afe,i}| }{2}  \right)^2, 
\end{equation}
where we have symmetrized the contribution of each element by taking the average of the positive and negative uncertainties. 

\begin{figure}[!htb]
  \centering
  \includegraphics[width=\columnwidth]{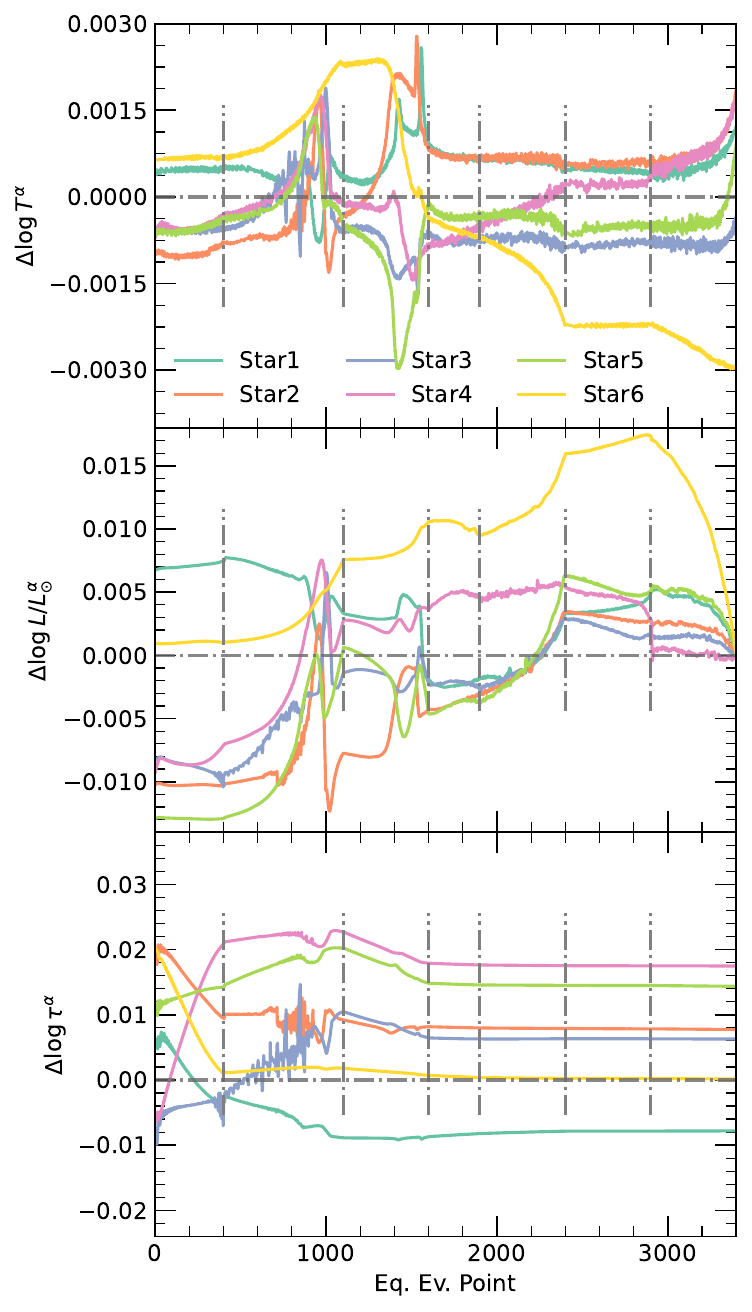}
  \caption{Same as Fig.~\ref{f:eeps_stars_syn} but comparing $\aevol$ and synthetic tracks.
  \label{f:eeps_stars_alpha} }
\end{figure}

The resulting $\aevol$ and errors for the six test cases are reported in Table~\ref{t:stars}. Figure~\ref{fig:tracks_stars} shows in red lines the evolutionary tracks obtained using a unique $\aevol$ enhancement for all \aelem. Results are almost indistinguishable from the full and the synthetic tracks, with only a small deviation seen close to the RGB-tip in the most extreme case, star 6, which shows large variations in the enhancement of its \aelem\ abundances. Figure~\ref{f:eeps_stars_alpha} shows, as a function of EEPs, the $\aevol$ tracks against the full evolutionary tracks. Differences in effective temperatures are always well below 1\% and for luminosities below 2\%, whereas for the age, they remain below 4\% in almost all cases except for Star 6, which shows an initial fractional difference reaching over 5\% in the initial evolutionary phase. It is relevant to mention that this type of detailed comparison between tracks depends on the specific definition of the main EEPs and the metric used to define all EEPs, but that the overall quantitative comparison is independent of this choice.

The small irregularities visible in the age differences in Figures \ref{f:eeps_stars_syn} and \ref{f:eeps_stars_alpha}, particularly for stars 2 and 3, may be related to the finite model resolution and interpolation effects near the main-sequence turnoff, where the transition between models with and without convective cores makes the interpolation of stellar tracks more challenging. Nevertheless, these effects are small, and the overall differences between the original and reconstructed tracks remain negligible for the purposes of this work.

Given the relevance of low-mass, $\alpha$-enhanced stars in Galactic Archaeology, we also assessed the robustness of our methodology by repeating the analysis for all six benchmark chemical patterns assuming a stellar mass of 0.80 $M_{\odot}$. This test was motivated by the fact that many old Galactic populations are composed of stars with masses below 1.00 $M_{\odot}$. We find that the inferred evolutionary $\aevol$ ratios remain essentially unchanged with respect to those obtained using the masses adopted in \ref{t:stars}, demonstrating that the determination of the evolutionary [$\alpha$/Fe] is not significantly affected by the adoption of a lower stellar mass.

\subsection{Relationship between $\aevol$, [Fe/H] and stellar mass}\label{ss:dependences}

The method presented to compute $\aevol$ relies on a reference track that, so far, we have chosen to be of the same $\feh$ and mass as the target star. The prior need of $\feh$ is not a limitation for the method, as this quantity is readily available in any spectroscopic survey. The stellar mass, on the other hand, is almost always unknown, and it is, in fact, one of the fundamental stellar parameters to be determined from observations. Here, we study the dependence of the $\aevol$ determination as a function of the mass of the reference track and show that it is minimal, i.e. that for all practical effects it can be considered independent of mass. This renders our method of general applicability to any spectroscopic survey, as it does not require additional prior information about the star.

\begin{figure*}
  \centering
  \includegraphics[width=\textwidth]{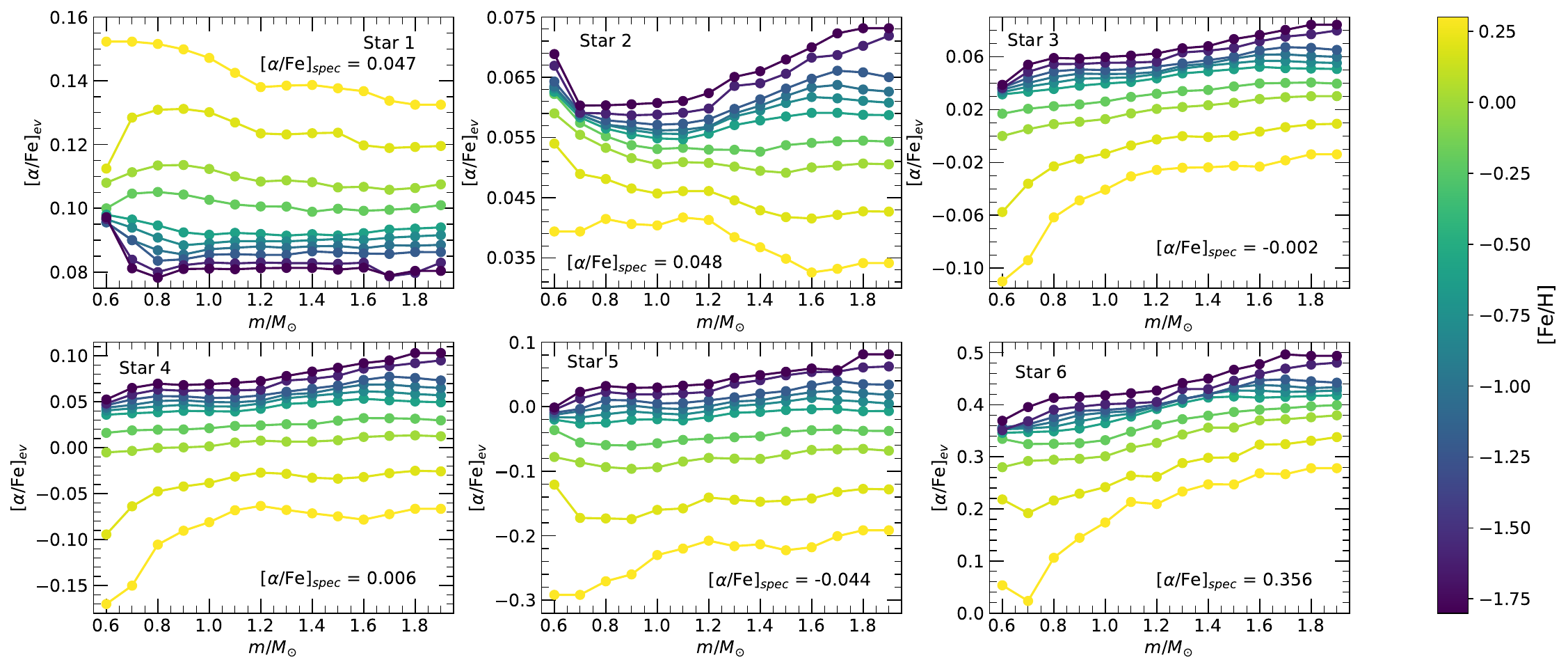}
  \caption{For stars 1 to 6, $\aevol$ is shown as a function of the mass of the reference track and for different [Fe/H] values as indicated in the figure.
  \label{f:alpha_bf_mass} }
\end{figure*} 

\begin{figure*}
  \centering
  \includegraphics[width=\textwidth]{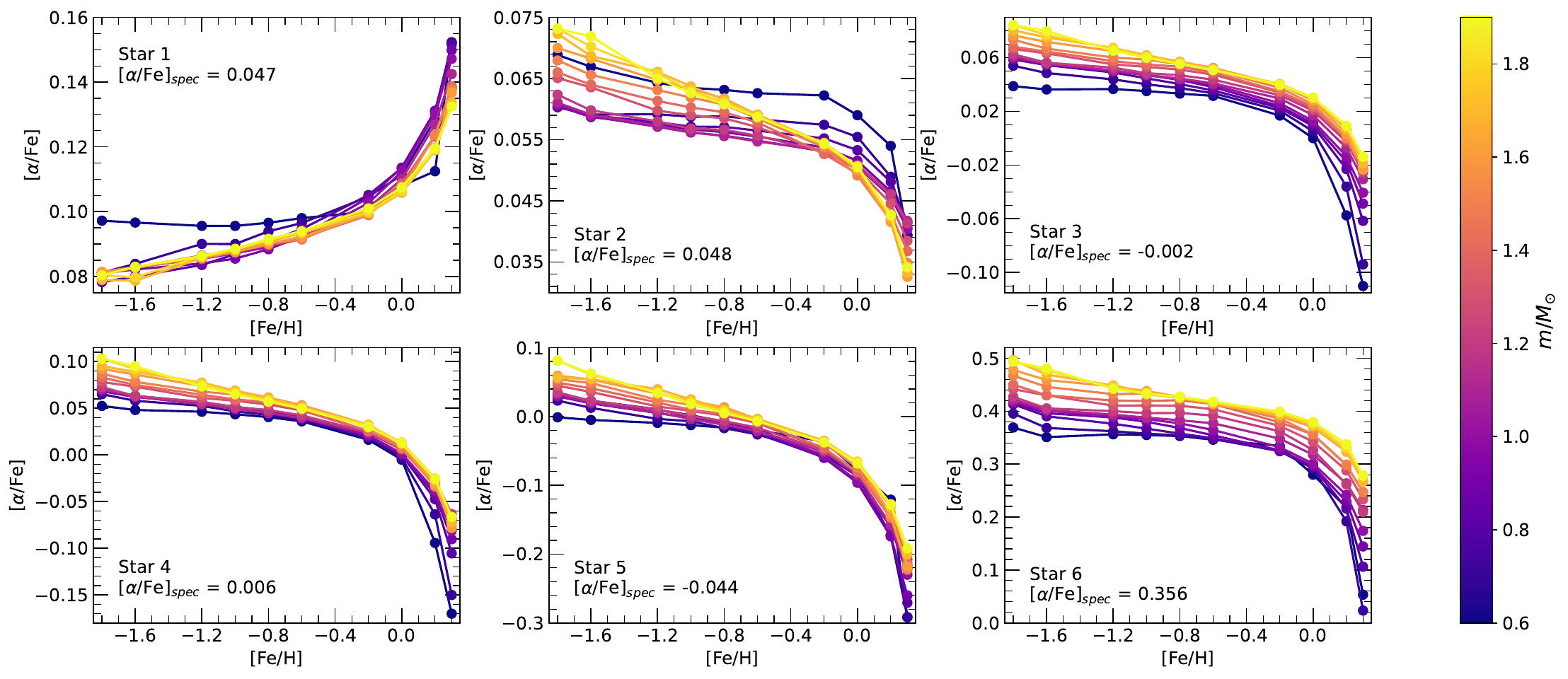}
 \caption{Same as Fig.~\ref{f:alpha_bf_mass} as a function of $\feh$ for different masses.}
  \label{f:alpha_bf_feh}
\end{figure*}

To show this, we have computed $\aevol$ for the six test stars using a wide range of reference tracks, varying both mass and $\feh$. Results are presented in Figure~\ref{f:alpha_bf_mass}, in which $\aevol$ is shown as a function of the reference mass for different $\feh$ values. For each star, it can be seen that the mass dependence is small at fixed $\feh$. Moreover, the behavior of $\aevol$ with mass is typically monotonic, with very few exceptions. On the other hand, it is clear from these plots that the $\feh$ value of the reference track has a strong impact on $\aevol$. Panels in Figure~\ref{f:alpha_bf_feh} show the same results but as a function of $\feh$. Here, at each $\feh$ value, the small scatter represents the dependence of $\aevol$ on the mass of the reference track. 

\begin{figure}[!htb]
  \centering
  \includegraphics[width=\columnwidth]{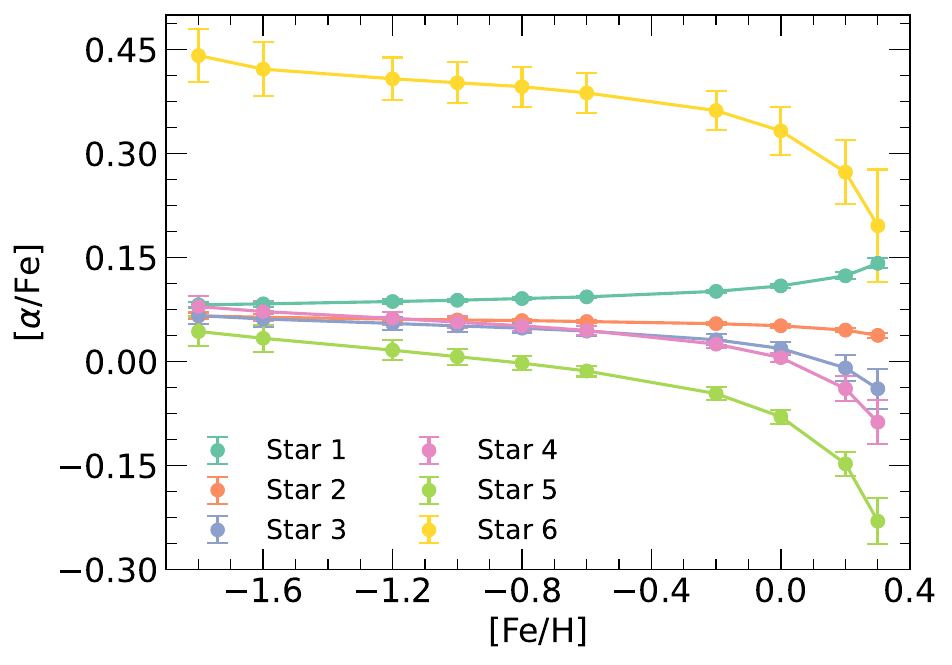}
  \caption{Average and standard deviation (shown as error bars) of $\aevol$ over stellar mass, as a function of $\feh$ reference track values for our test stars.}
  \label{fig:alpha_feh}
\end{figure}

The fundamental result of this section is presented in Figure~\ref{fig:alpha_feh}, which shows for all our test cases the mean and standard deviation (dispersion) of the $\aevol$ values obtained for different reference masses.
The dispersion in $\aevol$ values due to the choice of reference mass is always small, and the choice of the mass of the reference track becomes unimportant. The plot also summarizes the dependence of $\aevol$ on $\feh$. 
We note that, in a real life situation, only the $\aevol$ corresponding to the true observed $\feh$ would be required for each case.

\section{$\aevol$ tool calculator: \texttt{ALCHEMY}}\label{s:alchemy}


The methodology presented in previous sections allows us to exploit the rich information provided by large-scale spectroscopic surveys while, at the same time, taking advantage of the relative simplicity of using available libraries of stellar models and isochrones in which the treatment of chemical abundances is limited to $\feh$ and $\afe$. 

To streamline the calculation of $\aevol$ and its uncertainty for any \aelem\ pattern, we implemented the method described in Section \ref{ss:synthetic_alpha} in \texttt{ALCHEMY} (ALpha CHEmical abundances Modeling sYstem), a Python-based tool  publicly available on Zenodo\footnote{Available at: \href{https://zenodo.org/records/19664666}{https://zenodo.org/records/19664666}}.

Supported by the discussion in Sect.~\ref{ss:dependences}, \texttt{ALCHEMY} uses only coefficient tables obtained from evolutionary tracks 1.00 $M_{\odot}$. The procedure to compute $\aevol$ is rather simple:
\begin{enumerate}
\item coefficients $\abcoeff$ of 1.00 $M_{\odot}$ tracks are interpolated to the observed $\feh$,
\item a synthetic 1.00 $M_{\odot}$ track is computed for the observed chemical pattern as in Sect.~\ref{ss:synthetic_linear};
\item the same coefficients are used to build the residuals and the penalty function $F$ (Eq.~\ref{eq:epsilon} and ~\ref{eq:f_alpha}) using the spectroscopic \aelem\ pattern;
\item $\aevol$ and $\sigma_{\aevol}$ are computed as in Sect.~\ref{ss:synthetic_alpha}, the latter using the spectroscopic uncertainties of \aelem\ abundances. 
\end{enumerate}

Table~\ref{t:stars} lists, in its last row, the $\aevol$ values obtained using \texttt{ALCHEMY}, i.e. using $\abcoeff$ restricted to 1.00 $M_{\odot}$ tracks. The comparison with $\aevol$ obtained using the full set of $\abcoeff$ coefficients are very small, in all cases much smaller than $\sigma_{\aevol}$.

\section{Results} \label{s:results}

\subsection{$\aevol$ calculation for large-scale surveys} 

To illustrate the differences between spectroscopic $\afe$ and $\aevol$, we have applied our method to compute $\aevol$ for a large sample of stars in APOGEE, GALAH, and LAMOST. In this context, it is important to notice that each survey provides abundances for different chemical elements and the methodology of computing and reporting $\afe$ also varies. 

From APOGEE, we focus on stars in the APOKASC-3 red giants catalog \citep{2025ApJS..276...69P} and use  spectroscopic abundances from DR17 \citep{2022ApJS..259...35A}, for which elemental abundances for O, Mg, Si, S, Ca, and Ti are provided. The spectroscopic $\afe$, however, is not directly measured from a specific set of \aelem\ lines, but instead emerges from the global spectral fitting performed by the ASPCAP pipeline \citep{2016AJ....151..144G} in which [$\alpha$/M] is treated as one of the free parameters alongside $\teff$, $\log g$, and [M/H] (here M represents the total metallicty). This [$\alpha$/M] parameter represents a single dimension in the model grid that scales several $\alpha$-elements (primarily O, Mg, Si, S, Ca, and Ti) together, effectively encoding a composite $\alpha$-enhancement that is weighted by the spectral sensitivity of these elements in different stellar regimes. The final spectroscopic [$\alpha$/Fe] value is then obtained by expressing this global $\alpha$-enhancement relative to iron, using the Fe abundance derived from the same fit or from dedicated Fe lines, so that in practice [$\alpha$/Fe] $\simeq$ [$\alpha$/M] - [Fe/M]. As a result, the APOGEE [$\alpha$/Fe] should be interpreted as a model-dependent, spectrally weighted combination of multiple $\alpha$-elements rather than a simple average of individually measured abundances.

In the case of LAMOST, we have used results from DR10. The available set of elements is the same as in APOGEE. We have used the $\afe$ values derived from the LASP pipeline. This is obtained using convolutional neural networks (CNN). In this approach, a CNN is trained on a set of stars that have reliable labels, typically from high-resolution surveys such as APOGEE, so that the network learns the relationship between the low-resolution LAMOST spectra (R$\sim$1800) and stellar parameters, including $\afe$. The CNN automatically extracts features from the spectrum (such as broad absorption regions sensitive to Mg, Ca, or Ti) through its convolutional layers, without explicitly selecting spectral lines, and maps these patterns to the corresponding $\alpha$-enhancement values. Once trained, the network predicts $\afe$ for new LAMOST stars by recognizing similar spectral signatures. As a result, the inferred $\afe$ is a data-driven estimate whose scale and meaning are anchored to the training set, effectively making it a learned proxy for global $\alpha$-enhancement rather than a direct abundance determination.

For GALAH, we have used the DR4 results, in which the $\afe$ ratios are computed as the average of the individual [Mg/Fe], [Si/Fe], [Ca/Fe], and [Ti/Fe] abundances. GALAH DR4 also reports [O/Fe], although it is not used in their calculation of $\afe$ (Sven Buder, private comm.).

In order to compute $\aevol$\ with \texttt{ALCHEMY}, we have used the APOKASC-3 sample of red giants, and for LAMOST and GALAH we have used stars in the parameter space defined by 3500 $\leq T_{\text{eff}} \leq$ 5500, 1.00 $\leq \log g \leq$ 3.50 and -2.60 $\leq$ [Fe/H] $\leq$ 0.60. Regarding chemical elements, we have used those listed above for each survey. For GALAH, we have assumed that ${\rm [S/Fe]=[Si/Fe]}$ and, for all surveys, ${\rm [Ne/Fe]=[O/Fe]}$ because neon cannot be measured spectroscopically in cool stars. 

\begin{figure}[!htb]
  \centering
  \includegraphics[width=\columnwidth]{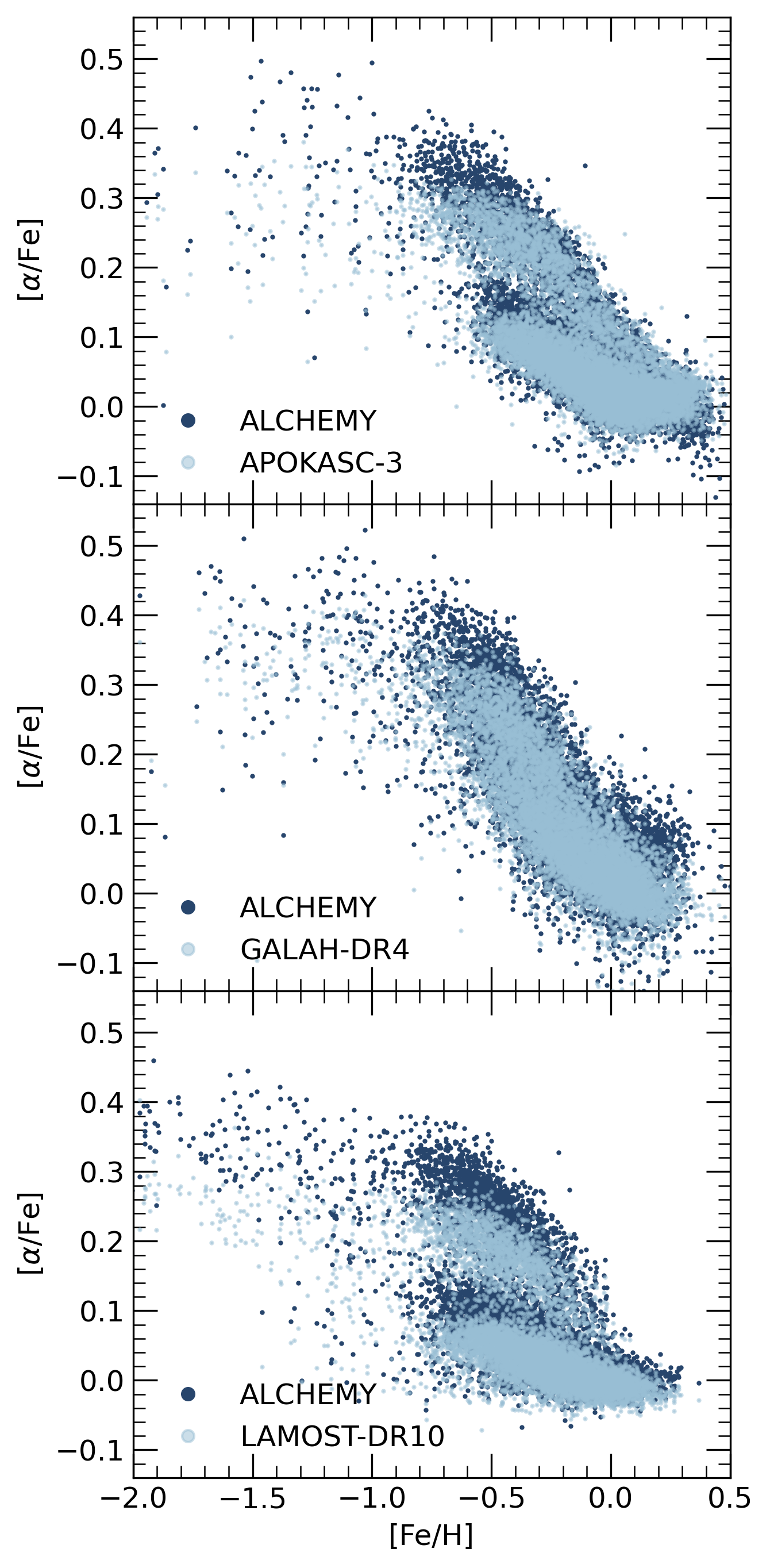}
  \caption{Tinsley-Wallerstein diagram comparing $\aevol$  with the $\afe$ values reported by the respective surveys.}
  \label{tinsley-wallerstein diagram}
\end{figure}

Figure~\ref{tinsley-wallerstein diagram} presents the Tinsley-Wallerstein diagram for the three surveys, showing both the spectroscopically derived and our $\aevol$ for the three surveys. As expected, for low $\afe$ values, results are quite similar in the comparison for all surveys. However, as $\afe$ increases towards lower $\feh$ values, $\aevol$ values tend to be larger than the spectroscopic ones. This implies that using the spectroscopic $\afe$ values will underestimate the impact that detailed chemical patterns have on the evolution of stars. Besides the general trend already described, the results for each survey show different behavior. In the case of APOGEE, results retain the structure from the spectroscopic results but stretch $\aevol$ in the low $\feh < -0.5$ region. For LAMOST, however, the differences more closely resemble a systematic offset across the entire $\feh$ range. We find the largest differences in the case of GALAH, for which the slope of $\afe$ vs $\feh$ is smaller than in the other two surveys. 

The comparison of results across these surveys clearly shows that spectroscopic $\afe$ should not be used directly in stellar evolution. Not only is the spectroscopic $\afe$ not the one most relevant for stellar evolution, but every survey needs to be corrected differently. The methodology we present in this paper corrects for these deficiencies.

\subsection{Minimum set of chemical elements and handling of incomplete information} \label{ss:minimal}

\begin{figure*}
  \centering
  \includegraphics[width=\textwidth]{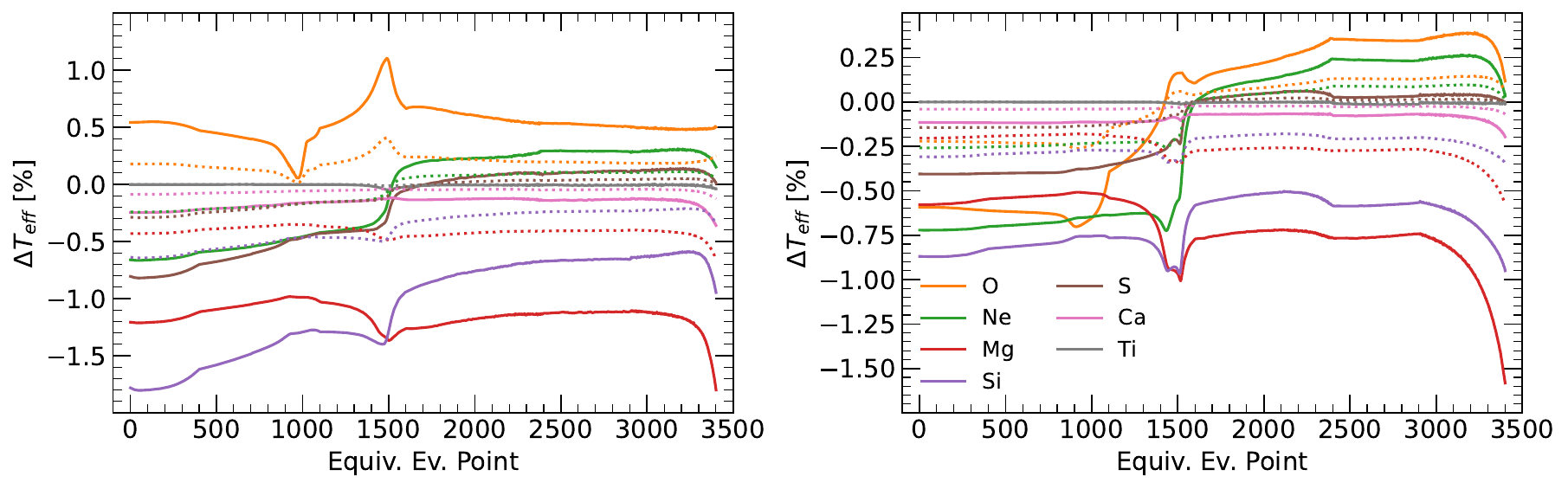}
 \caption{Difference in effective temperature for 1~\msun\ evolutionary tracks induced by changes in \aelem\ abundances. Left and right panels show results for $\feh=0.00$ and $-0.60$ respectively, and dotted and solid lines depict enhancements of 0.10 and 0.25 respectively in each element. \label{fig:minimalset}}
\end{figure*} 

It is not possible to determine spectroscopically the complete pattern of \aelem\ abundances in cool stars. The most obvious drawback is the impossibility of measuring neon at all in them. The wavelength coverage of spectrographs impose a technical, not physical, restriction, as the lack of sulfur in GALAH measurements illustrates. The question then arises: which is the minimal set of \aelem\ required to determine an $\aevol$ that is appropriate for stellar evolution models? This question is related to our detailed discussion in Sect.~\ref{ss:cmpeffect}. Here, we take a more pragmatic approach focused only on $\aevol$ calculations. 

In Fig.~\ref{fig:minimalset}, we show the variation in $\teff$ induced by changes in the abundance of \aelem\ along 1~\msun\ evolutionary tracks of $\feh=0.00$ and $-0.60$ used to compute $\aevol$ (Sect.~\ref{s:alchemy}). Roughly, the transition from main sequence/subgiant phases towards the red giant branch is located around the EEP 1500. 
The detailed morphology of temperature differences shown in Fig.~\ref{fig:minimalset} depends on the specific choice of main EEPs (see appendix). However, the results discussed below are of general validity.  

Of all elements, Ca and Ti have a negligible role. A spectroscopic determination of $\afe$ that relies to some extent on these elements is not a good choice for stellar evolution models. On the contrary, Mg and Si are the most dominant elements and, fortunately, are almost always available. Oxygen is also needed, as it strongly affects not only opacities, but the initial helium of models if this is computed from the total metallicity as we have done (Eq.~\ref{eq:y-z}). This is a customarily choice in many libraries of stellar evolution models and isochrones available in the literature. Unfortunately, O is a difficult element to determine reliably in cool stars spectra.  In such cases, an assumption must be made about its enhancement, relating it to Mg and/or Si enhancements, guided by data or chemical evolution models. Something similar happens with Ne, and we recommend that it always be included in the calculation of $\aevol$, with the assumption that its enhancement matches that of O. Finally, sulfur has an impact on the main sequence, particularly around solar-like metallicities, and less so at lower $\feh$ values. On the red giant branch it is not relevant. 

In summary, in order of relevance, Mg, Si, O, Ne (linked to O) are needed to have an accurate calculation of $\aevol$, and S is desirable particularly if stars are main sequence/subgiants, but it is only of minor relevance for red giants.

\section{Summary and Conclusions}\label{s:conclusion}

We have developed a method to obtain stellar evolution tracks for general \aelem\ abundance patterns without the need to carry out detailed stellar evolutionary calculations. The approach is based on linearly combining the effects induced by individual chemical elements, whose contributions can in turn be described as quadratic perturbations relative to reference evolutionary tracks with solar-scaled abundances. These so-called synthetic evolutionary tracks reproduce full stellar evolution calculations with high fidelity.

As a second step, we devise a method to determine a single $\alpha$-enhancement value for any observed $\alpha$-element pattern. This evolutionary $\aevol$ is defined as the constant (across elements) $\alpha$-enhancement that best reproduces the evolution of a star with the observed $\alpha$-pattern. In doing this, we recover the ability to use standard grids of stellar models and isochrones usually built with $\feh$ and $\afe$ as the only two dimensions defining the chemical composition of stars. Importantly, our method allows mapping spectroscopic observations into an $\alpha$-enhancement that is meaningful for stellar evolution, away from the particular idiosyncrasies of spectroscopic surveys when reporting $\alpha$-enhancements. In order to apply our method to large-scale spectroscopic surveys, we have developed \texttt{ALCHEMY}, a simple numerical code that allows the user to compute $\aevol$ and its uncertainty for any set of observed \aelem\ pattern. $\aevol$ can then be used to determine stellar parameters in combination with any available grid of stellar evolution models or isochrones. 

We have shown that the difference between $\aevol$ and $\afe$ is survey-dependent. This is expected, as spectroscopic surveys determine and report $\afe$ in different ways, and the spectroscopic $\afe$ is also different from $\aevol$, the relevant quantity for stellar evolution. For the first time, with \texttt{ALCHEMY}, we can translate detailed spectroscopic abundance determinations for \aelem\ into the correct value needed to compute stellar evolution models.

Future work includes the application of our newly developed method to redetermine fundamental stellar parameters of APOKASC-3 red giants catalog and to extend the methodology to include, in a similar way, variations of C and N, the two non-\aelem\ relevant for stellar evolution. 

\begin{acknowledgements}
This work was supported by grant PID2023-149918NB-I00 from the Spanish Ministerio de Ciencia, Innovaci\'on y Universidades (MCIU), the Spanish program Unidad de Excelencia Mar\'ia de Maeztu CEX2020-001058-M and by the MaX-CSIC Excellence Award MaX4-SOMMA-ICE. PDR work has been carried  out within the framework of the doctoral program in Physics of the Universitat Aut\`onoma de Barcelona.
\end{acknowledgements}



%
%

\bibliographystyle{aa}
\bibliography{references}

\appendix

\section{Normalized stellar evolution models}\label{appendix a}

As described in section~\ref{ss:models}, once evolutionary tracks are interpolated into a normalized set of equivalent evolutionary points (EEPs) common to the complete grid of evolutionary tracks. This is done by defining a set of main EEPs as described below, and an appropriate metric between them. The main EEPs ($x^p$) are:
\begin{enumerate}
    \item \textbf{Zero-Age Main Sequence:} The stage where the central hydrogen abundance decreases slightly, reaching $X_c$ = 0.9997 $X_{c,0}$, where $X_{c,0}$ is the initial hydrogen mass fraction.
    \item \textbf{Main Sequence:} Identified as the point just before the central hydrogen abundance drops below 60$\%$ of its initial value, i.e. $X_c \geq$ 0.60 $X_{c,0}$.
    \item \textbf{Main Sequence to Sub-giant Branch Transition:} The moment when all central hydrogen is exhausted, i.e. $X_c$ = 0.00.
    \item \textbf{Base of the Red Giant Branch:} The beginning of the Red Giant Branch phase, marking the transition from the sub-giant branch.
    \item \textbf{Red Giant Branch luminosity increase:} The point where the luminosity has increased by 0.3 dex compared to the previous point.
    \item \textbf{Red Giant Branch bump (1):} The stage where the luminosity reaches a local maximum and the temperature reaches a local minimum, satisfying $\log L_{i} - \log L_{i + 1} \geq$ 0.00 and  $\log T_{i} - \log T_{i + 1} \leq$ 0.00.
    \item \textbf{Red Giant Branch bump (2):} The point where the luminosity reaches a local minimum, and the temperature reaches a local maximum, satisfying $\log L_{i} - \log L_{i + 1} \leq$ 0.00 and  $\log T_{i} - \log T_{i + 1} \geq$ 0.00.
    \item \textbf{Red Giant Branch luminosity threshold:} The point where the luminosity reaches a value $\log L \leq$ 3.10, after the Red Giant Branch Bump.
\end{enumerate}


In between each pair of main EEPs, we define a set of secondary EEPs ($x^s$). These are equally spaced points according to the following  metric:
\begin{equation}\label{metric}
    x^s_i = x^s_{i-1} + \frac{|X_{c,i} - X_{c,i-1}|}{X_c^{max} - X_c^{min}}, \quad \text{$\forall$ $x^p$ < 3}
\end{equation}
\begin{equation}
    x^s_i = x^s_{i-1} + \frac{|\tau_{i} - \tau_{i-1}|}{\tau_{max} - \tau_{min}} + \frac{1}{5} \left( \frac{|\log L/L_{\odot,i} - \log L/L_{\odot,i-1}|}{\log L/L_{\odot,max} - \log L/L_{\odot,min}} \right)
\end{equation}

Here, $X_c$ is the central hydrogen mass fraction, $\tau$ is the age, and $L$ the luminosity.

The normalization of the tracks sets the same number of points for each model. All models have a total of 3400 points (including the main EEPs) and are distributed as follows: 400 points between EEP 1 and 2, 700 between EEP 2 and 3, 500 for EEPs 3 to 4, 300 from EEP 4 to EEP 5 and 500 points between each pair in the range 5 to 8. 

\begin{figure}[!htb]
  \centering
  \includegraphics[width=\columnwidth]{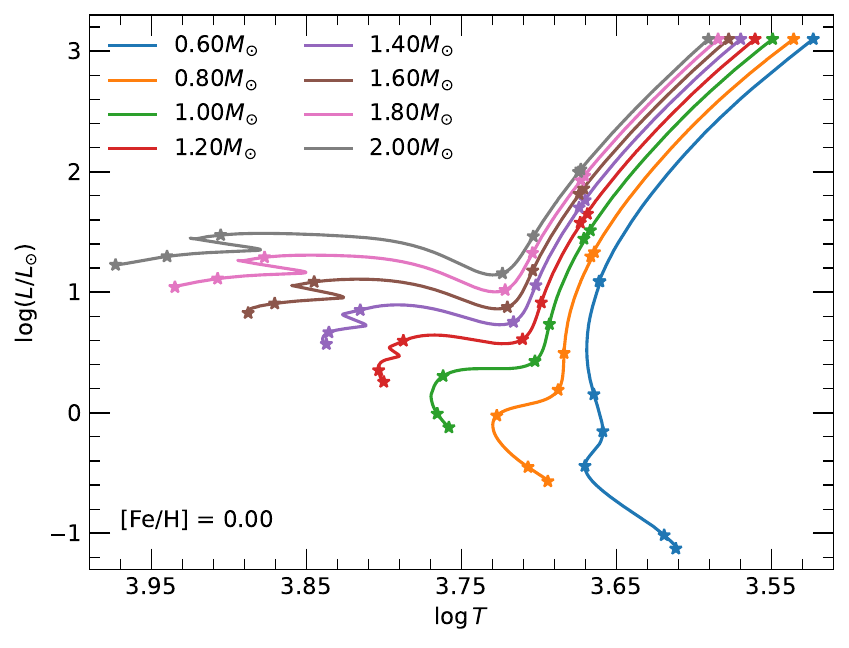}
  \caption{Distribution of the eight main EEPs along the zero-age main sequence, sub-giant branch and red giant branch phases of stellar evolutionary models with different stellar masses and [Fe/H] = 0.00.}
  \label{f:eeps_hrd}
\end{figure}

Figure~\ref{f:eeps_hrd} shows the location of the main EEPs for several tracks of different stellar masses.

\end{document}